\documentclass[prl,showpacs,showkeys,twocolumn,notitlepage,10pt,superscriptaddress]{revtex4-1}
\usepackage{amsfonts,bbold,amsmath,amssymb,graphicx,epstopdf,verbatim,dsfont,color,hyperref}
\usepackage[english]{babel}

\newcommand{\tr}{{\rm Tr}}

\def \be{\begin{equation*}}
\def \ee{\end{equation*}}

\begin{document}

\title{Quantum approximate Bayesian computation for NMR model inference}
\author{Dries Sels}
\affiliation{Department of Physics, Harvard University, 17 Oxford st., Cambridge, MA 02138, USA}
\affiliation{Theory of quantum and complex systems, Universiteit Antwerpen, B-2610 Antwerpen, Belgium}
\author{Hesam Dashti}
\affiliation{Division of Preventive Medicine, Brigham and Women’s Hospital, Harvard Medical School, 900 Commonwealth Ave., Boston, MA 02215, USA}
\author{Samia Mora}
\affiliation{Division of Preventive Medicine, Brigham and Women’s Hospital, Harvard Medical School, 900 Commonwealth Ave., Boston, MA 02215, USA}
\affiliation{Division of Cardiovascular Medicine, Brigham and Women’s Hospital, Harvard Medical School, 900 Commonwealth Ave., Boston, MA 02215, USA}
\author{Olga Demler}
\affiliation{Division of Preventive Medicine, Brigham and Women’s Hospital, Harvard Medical School, 900 Commonwealth Ave., Boston, MA 02215, USA}
\author{Eugene Demler}
\affiliation{Department of Physics, Harvard University, 17 Oxford st., Cambridge, MA 02138, USA}
\date{\today}

\begin{abstract}
Recent technological advances may lead to the development of small scale quantum computers capable of solving problems that cannot be tackled with classical computers. A limited number of algorithms has been proposed and their relevance to real world problems is a subject of active investigation. Analysis of many-body quantum system is particularly challenging for classical computers due to the exponential scaling of Hilbert space dimension with the number of particles. Hence, solving problems relevant to chemistry and condensed matter physics are expected to be the first successful applications of quantum computers.  In this paper, we propose another class of problems from the quantum realm that can be solved efficiently on quantum computers: model inference for nuclear magnetic resonance (NMR) spectroscopy, which is important for biological and medical research.
Our results are based on three interconnected studies. Firstly, we use methods from classical machine learning to analyze a dataset of NMR spectra of small molecules. We perform a stochastic neighborhood embedding and identify clusters of spectra, and demonstrate that these clusters are correlated with the covalent structure of the molecules. Secondly, we propose a simple and efficient method, aided by a quantum simulator, to extract the NMR spectrum of any hypothetical molecule described by a parametric Heisenberg model. Thirdly, we propose a simple variational Bayesian inference procedure for extracting Hamiltonian parameters of experimentally relevant NMR spectra.
\end{abstract}
\maketitle


\section*{Introduction}
One of the central challenges for quantum technologies during the last few years has been a search for useful applications of near-term quantum machines~\cite{preskill18}. While considerable progress has been achieved in increasing the number of qubits and improving their quality~\cite{bernien17,friis18}, in the near future, we expect the number of reliable gates to be limited by noise and decoherence; the so called Noisy Intermediate-Scale Quantum (NISQ) era. As such, hybrid quantum-classical methods have been proposed to make the best out of the available quantum hardware and supplement it with classical computation. Most notably, there has been the development of the Quantum Approximate Optimization Algorithm (QAOA)~\cite{farhi14} and the Variational Quantum Eigensolver (VQE)~\cite{peruzzo14}. Both algorithms use the quantum computer to prepare variational states, some of which might be inaccessible through classical computation, but use a classical computer to update the variational parameters. A number of experiments have already been performed, demonstrating the feasibility of these algorithms~\cite{kokail19,kandala17,colless18}, yet their bearing on real world problems remains unclear.
In model-based statistical inference one is often faced with similar problems. For simple models one can find the likelihood and maximize it but for complex models the likelihood is typically intractable~\cite{diggle84,beaumont02}. Nuclear magnetic resonance (NMR) spectroscopy is a perfect example: there is a good understanding of the type of model that should be used (see equation (1))  and one only needs to determine the appropriate parameters. However, computing the NMR spectrum for a specific model requires performing computations in the exponentially large Hilbert space, which makes it extremely challenging for classical computers. This feature has been one of the original motivations for proposing NMR as a platform for quantum computing~\cite{gershenfeld350}. While it has been shown that no entanglement is present during NMR experiments~\cite{braunstein99,menicucci02}, strong correlations make it classically intractable, i.e. the operator Schmidt rank grows exponentially which for example prohibits efficient representation through tensor networks~\cite{datta07}
 Its computational power is between classical computation and
deterministic quantum computation with pure states~\cite{knill98}, which makes it an ideal candidate for hybrid quantum-classical methods. As we argue below, the required initial quantum states can be prepared by low depth circuits and the problem is robuust against decoherence. By simulating the model on a quantum computer, it runs efficiently while the remaining inference part is simply solved on a classical computer. One can think of this as an example of quantum Approximate Bayesian Computation (qABC), putting it in the broader scope of quantum machine learning methods~\cite{biamonte17}. In contrast to most of the proposed quantum machine learning applications, the present algorithm does not require challenging routines such as amplitude amplification~\cite{brassard,grover} or Harrow-Hassidim-Lloyd (HHL) algorithm~\cite{hhl}.
\section*{NMR-spectroscopy}
NMR spectroscopy is a spectroscopic technique which is sensitive to local magnetic fields around atomic nuclei. Typically, samples are placed in a high magnetic field while driving RF-transitions between the nuclear magnetic states of the system. Since these transitions are affected by the intramolecular magnetic fields around the atom and the interaction between the different nuclear spins, one can infer details about the electronic and thus chemical structure of a molecule in this way. One of the main advantages of NMR is that it is non-destructive, in contrast to, for example, X-ray crystallography or mass spectrometry. This makes NMR one of the most powerful analytical techniques available to biology~\cite{bothwell11}, as it is suited for in vivo and in vitro studies~\cite{jonghee19}. NMR can, for example, be used for identifying and quantifying small molecules in biological samples (serum, cerebral fluid, etc.)~\cite{olaf07,larive15,napolitano13}. On the other hand, NMR experiments have limited spectral resolution and as such face the challenge of interpreting the data, since extracted information is quite convoluted. We only directly observe the magnetic spectrum of a biological sample, whereas our goal is to learn the underlying microscopic Hamiltonian and ultimately identify and quantify the chemical compounds. While this inference is tractable for small molecules, it quickly becomes problematic, making inference a slow and error-prone procedure~\cite{ravanbakhsh15}. The analysis can be simplified by incorporating a priori spectral information in the parametric model~\cite{degraaf90}. For that purpose, considerable attention has been devoted to determining NMR model parameters for relevant metabolites such as those found in plasma, cerebrospinal fluid and mammalian brains~\cite{wevers94,wevers95,govindaraju00,hesam1,hesam2,chemshift,chemshiftAI}.

In what follows we will be concerned with 1D proton NMR but generalization to other situations are straightforward. For liquid $^1$H-NMR, a Heisenberg Hamiltonian 
\begin{equation}
H(\theta)=\sum_{i,j} J_{ij} \mathbf{S}_i \cdot \mathbf{S}_j+\sum_i h_i \mathbf{S}^x_i, 
\label{eq:H}
\end{equation}
yields a reasonable effective description for the nuclear spins, where $\theta$ explicitly denotes the dependence of the Hamiltonian on its parameters $\theta=\{J_{ij},h\}$. Here $J_{ij}$ encodes the interaction between the nuclear spins $\mathbf{S}$ and $h_i$ is the effective local magnetic field. Note that this Hamiltonian contains two essential approximations (i) the interactions are chosen to $SU(2)$ invariant and (ii) the local magnetic fields -- called chemical shifts in the NMR literature -- are unidirectional. The rationale for the latter is that most of these local magnetic fields are caused by diamagnetic screening due to electronic currents induced by the large external magnetic field. This field will tend to oppose the external field and hence be largely uniaxial. For liquid state NMR, the rapid tumbling of the molecules averages out the dipar coupling between the nuclei, approximately resulting in isotropic exchange interactions between nuclear spins~\cite{levitt}. The fact that the interactions are rotationally invariant, allows us to remove the average (external) field from the Hamiltonian, i.e. $\mathbf{S}^x_{tot}=\sum_i \mathbf{S}^x_i$ commutes with Hamiltonian~\eqref{eq:H} and will therefore only shift the NMR spectrum.

Within linear response the evolution of the system subject to a radio frequency $z$-magnetic field is determined by the response function:
\begin{equation}
S(t|\theta)= \tr\left[  e^{iH(\theta)t}\mathbf{S}^z_{tot}e^{-iH(\theta)t}\mathbf{S}^z_{tot}\rho_0\right],
\label{eq:Dt}
\end{equation}
where $\rho_0$ denotes the initial density matrix of the system and $\mathbf{S}^z_{tot}=\sum_i \mathbf{S}^z_{i}$. The measured spectrum is simply given by:
\begin{equation}
A(\omega|\theta)= {\rm Re} \int_0^\infty {dt} e^{i\omega t-\gamma t} S(t|\theta),
\label{eq:spectrum}
\end{equation}
where $\gamma$ is the effective decoherence rate. For room temperature $^1$H-NMR, the initial density matrix can be taken to be an infinite temperature state, i.e. 
\begin{equation}
\rho_0 \approx \frac{\mathbb{1}}{\tr \left[ \mathbb{1}\right] }.
\end{equation}
Indeed, even a 20 T magnetic field will only lead to a bare proton resonance frequency of about 900 MHz. In contrast, room temperature is about 40 THz, so for all practical purposes we can consider it equally likely for the spin to be in the excited state or in the ground state.  Chemical shifts $h_i$ are of the order of a few parts per million, resulting in local energy shifts of a few kHz, while the coupling or interaction strength $J$ is of the order of a few Hz. Despite these low frequencies and the high temperature of the system, one can typically still infer the parameters due to the small decoherence rate of the proton nuclear spin. Due to the absence of a magnetic quadrupole moment, the protons do not decohere from the electric dipole fluctuations caused by the surrounding water molecules. This gives the proton nuclear spin a coherence time of the order of seconds to tens of seconds, sufficiently long to create some correlations between the various spins. The remaining part of this work is concerned with the question of how to infer the model parameters $J_{ij}$ and $h_i$ of our effective Hamiltonian~\eqref{eq:H} from a measured spectrum~\eqref{eq:spectrum}.
\section*{Clustering}
\begin{figure}[t]
	\centering
	\includegraphics[width=0.23 \textwidth]{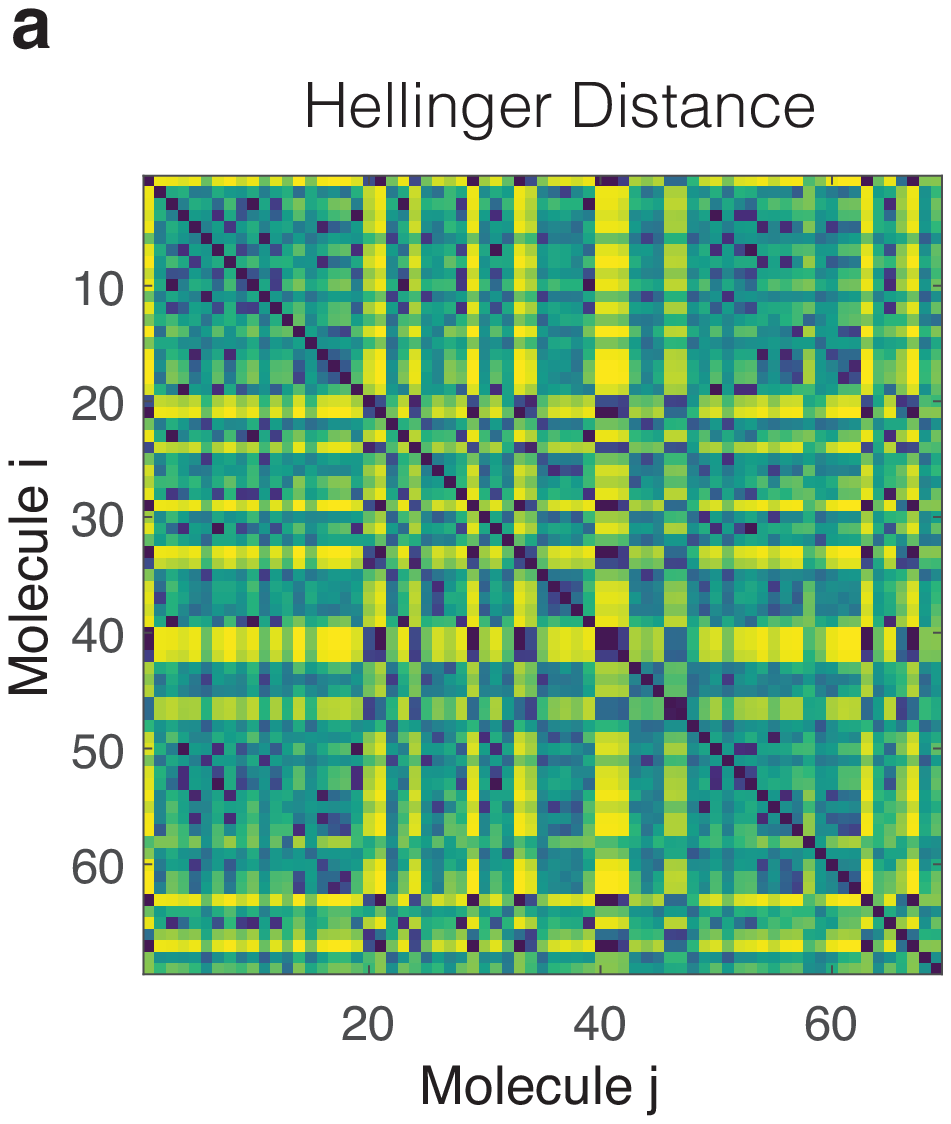}
	\includegraphics[width=0.23 \textwidth]{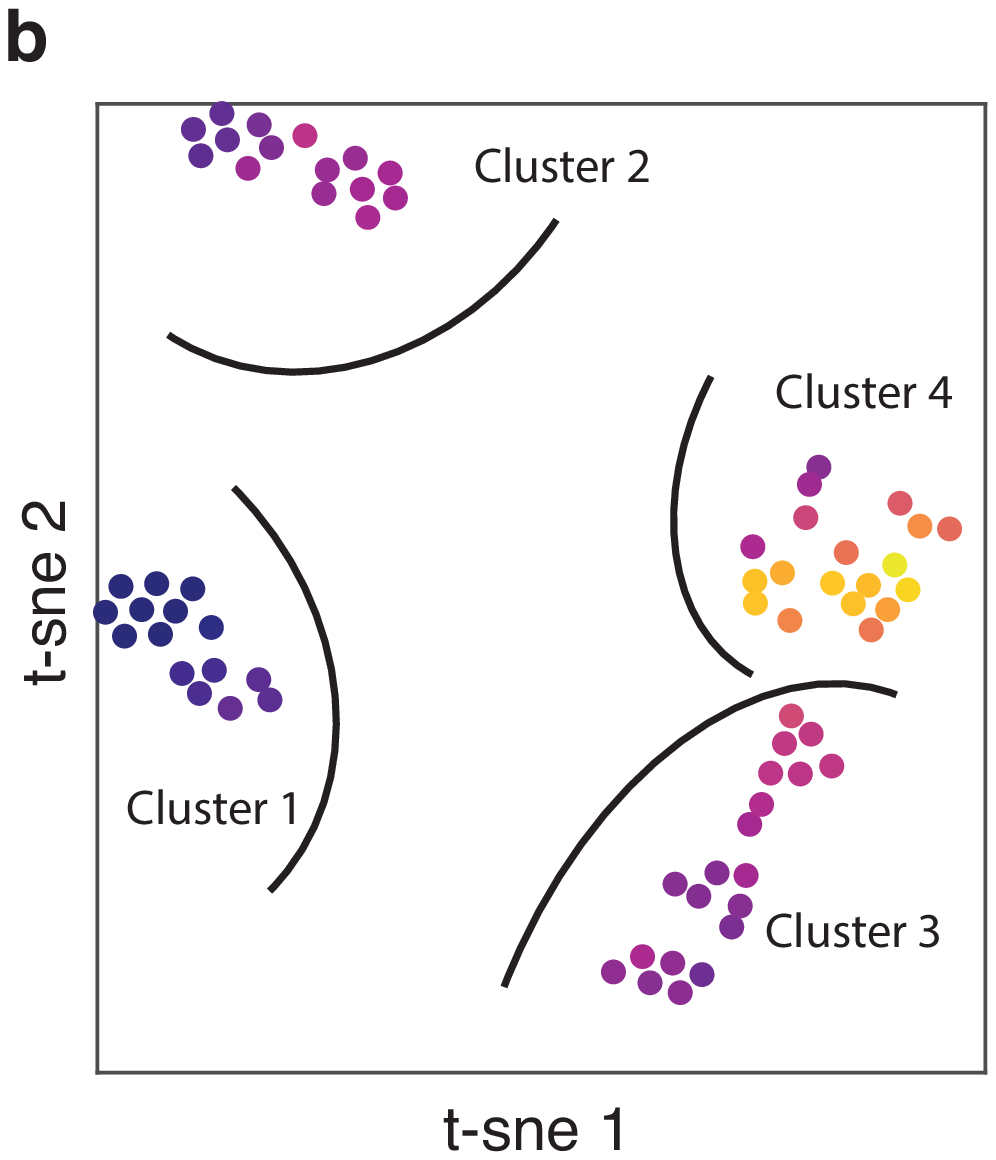}
	\caption{ \textbf{Clustering} In order to identify whether naturally occurring molecules have some atypical NMR spectrum, we perform a clustering analysis. In panel--A we show the distance between the various NMR spectra, where the Bhattacharyya coefficient is used to measure similarity. To obtain a meaningful comparison, spectra are shifted and scaled such that they are all centered around the same frequency and have the same bandwidth. To extract clusters we perform a t-SNE shown in panel--B with perplexity of 10; which is chosen because it has minimal Kullback-Leibler (KL) loss, i.e. the KL-loss was 0.145.}
	\label{fig:Distance}
\end{figure}

\begin{figure*}[t!]
	\centering
	\includegraphics[width=0.9 \textwidth]{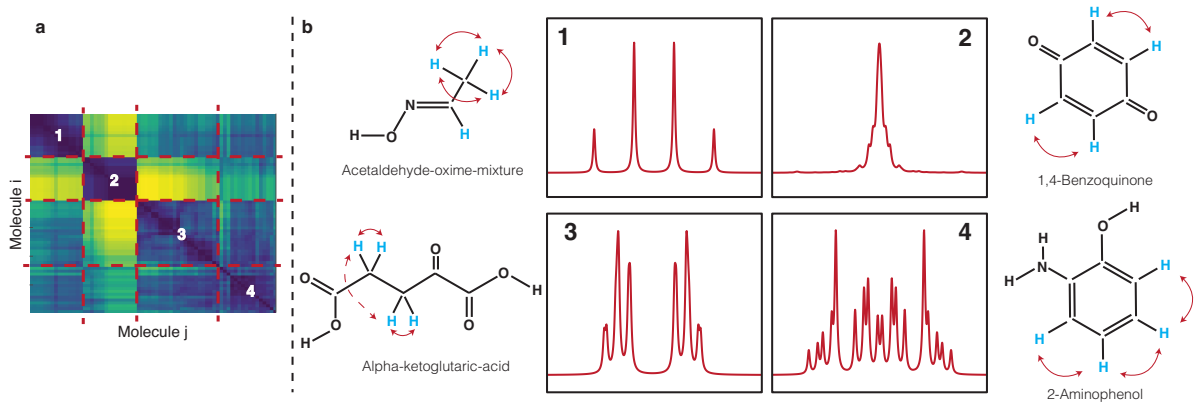}
	\caption{\textbf{NMR Spectra} By clustering the molecules according to the Hellinger distance t-SNE clusters we can reorganize the distance matrix as shown in panel A. For each of the clusters, we look at the different spectra, which indeed show great similarity. A representative spectrum for each of the clusters is shown in panel B, where the spectra are labeled according to the t-SNE clusters shown in Fig.~\ref{fig:Distance}--B. In addition, we show an example small molecule out of this cluster next to the associated spectrum. The atoms and interactions responsible for the shown portions of the spectra are indicated in blue and red arrows respectively.}
	\label{fig:ClusterSpectra}
\end{figure*}

Given real NMR data, summarized by the experimentally acquired spectrum $\mathcal{A}(\omega)$, our goal, in general, is to learn a parametrized generative model which explains how this NMR data is generated. Fortunately, we have a good idea about the physics which allows us to write down a model, i.e. expressions~\eqref{eq:spectrum}, that is close to reality thereby ensuring a small misspecification error. The drawback however is that the model is analytically intractable and becomes increasingly complex to simulate with increasing number of spins. In the next section we will discuss how to alleviate this problem by using a programmable quantum simulator to simulate the problem instead. Even if we can simulate our model~\eqref{eq:spectrum}, we still have to find a reliable and robust way to estimate the parameters $\theta$. Physical molecules have far from typical parameters $\theta$, see SI for a mathematical description. After all, if they do not, how could we infer any structural information out of the spectrum? 
To extract NMR spectral features, we first perform unsupervised learning on a dataset containing $69$ small organic molecules, all composed out of 4 $^1$H-atoms, observable in NMR 1D-$^1$H experiments. Their effective Hamiltonian parameters $\theta$ have previously been determined, which provides us with a labeled dataset to test our procedure. Furthermore, by only using the spectra themselves, we can use any relevant information as an initial prior for inference on unknown molecules. The dataset was compiled using the GISSMO library~\cite{hesam1,hesam2,hesam3}. In order to extract the structure in the dataset, we perform a $t$-distributed stochastic neighborhood embedding (t-SNE)~\cite{tsne08,lvdm19} to visualize the data in 2 dimensions. Fig~\ref{fig:Distance}--B, shows the 2-dimensional t-SNE embedding of the dataset based on the Hellinger distance shown in Fig.~\ref{fig:Distance}--A, a detailed comparison of different metrics is presented in the SI. The colorscale in panel B shows the inverse participation ratio of each sample, $
{\rm IPR}= \int_{-\infty}^\infty {\rm d}\omega A(\omega |\theta) / \int_{-\infty}^\infty  {\rm d}\omega A^2(\omega |\theta),
$, a measure for the total number of transitions that contribute to the spectrum. 
At least 4 well defined clusters are identified, density-based spatial clustering of applications with noise (DBSCAN)~\cite{dbscan} was used to perform the clustering. Using the clusters as indicated in Fig.~\ref{fig:Distance}--B, we can sort the molecules per cluster and have a look at the spectra. The sorted distance matrix is shown in Fig.~\ref{fig:ClusterSpectra}--A, it clearly shows we managed to find most of the structures in the system. In fact a closer look at the spectra of each of the clusters indeed reveals they are all very similar. Fig.~\ref{fig:ClusterSpectra}--B shows a representative spectrum for each of the clusters, as expected the IPR goes up if we go from cluster one to cluster four. All spectra in \emph{cluster 1} have the property of containing two large peaks and two small peaks, where the larger peak is about three times higher than the small peak. This is indicative of molecules with a methyl group (CH3) with its protons coupled with a methine proton (CH). One example of such structures can be seen in acetaldehyde oxime (BMRB ID~\cite{ulrich08}: bmse000467) (as shown to the left in Fig.~\ref{fig:ClusterSpectra}--B). The fact that the 3 protons are equivalent results in the 3:1 ratio of the peaks. Molecules from \emph{cluster 2} have are highly symmetric and have two pairs of two methine protons (CH) where the protons are on neighboring carbon atoms. The symmetry in the molecule makes the spectrum highly degenerate. In contrast, \emph{cluster 3} has molecules where there are two neighboring methylene groups (CH2). The interacting splitting causes a spectrum as shown in Fig.~\ref{fig:ClusterSpectra}--B. Finally, \emph{cluster 4} has four inequivalent protons with different chemical shifts and interactions between them. As a results, there is a plethora of possible transitions and the spectrum has an erratic form such as shown in Fig.~\ref{fig:ClusterSpectra}--B. In that sense, cluster 4 is most like a disordered quantum spin chain.

Given a new spectrum of an unknown molecule we can find out whether the molecule belongs to any of the identified molecular sub-structures, i.e. by computing the mean Hellinger distance to each of the identifed clusters one can robustly classify the spectra. In the supplemental information we present results in which we randomly choose 39 samples and consider those as clustered, while we use the other 30 samples to test the procedure. Samples that belong to cluster 1 and 2 are always correct classified. One sample from cluster 4 was misclassified for cluster 3.  Since we know the spin matrix $\theta_i$ for each of the molecules in the dataset, we have a rough estimate of what the Hamiltonian parameters are and where the protons are located with respect to each other. However, there is still a lot of fine structure within clusters, in particular in clusters 3 and 4, as can be seen in Fig.~\ref{fig:ClusterSpectra}--A. In what remains, we are concerned with finding an algorithm to further improve the Hamiltonian parameter estimation.

\section*{Quantum Computation}
\begin{figure}[h]
	\centering
	\includegraphics[width=0.45 \textwidth]{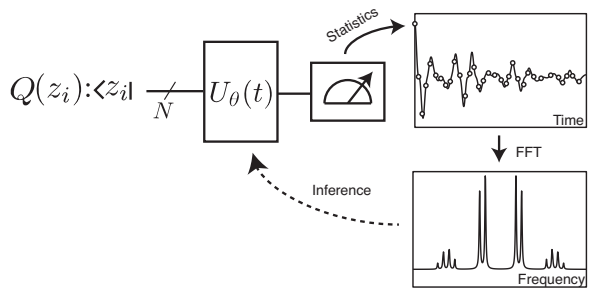}
	\caption{\textbf{Method overview} Take a product state $\left|z_i \right>$ with a given total magnetization $m_i$, according to  $Q_0(j)$. The latter can be choosen to minimze the variance of the estimand. After this initial preparation, we evolve the state under the Hamiltonian $H(\theta)$ and measure the project back onto the z-basis at time $t$. By applying a fast Fourier transform to the estimate $S(t|\theta)$, one obtains the spectrum which can be used to infer the parameters of the Hamiltonian.}
	\label{fig:Procedure}
\end{figure}

While our model is microscopically motivated, thereby capturing the spectra very well and allowing for a physical interpretation of the model parameters, it has the drawback that, unlike simple models such as Lorentzian mixture models~\cite{sokolenko19,ku19}, there is no analytic form for the spectrum in terms of the model parameters. Moreover, even simulating the model becomes increasingly complex when the number of spins increase. Before we solve the inference problem, let us present an efficient method to extract the simulated NMR spectrum on a quantum simulator-computer. The basic task is to extract the spectrum~\eqref{eq:spectrum} by measuring~\eqref{eq:Dt}. Recall that we work at infinite temperature, hence by inserting an eigenbasis of the total $z$-magnetization $\mathbf{S}^z_{tot}=\sum_j m_j \left |z_j\right> \left< z_j \right|$, we find
\begin{equation}
S(t|\theta)= \sum_{i,j} m_i m_j P_t(i|j,\theta)P_0(j),
\label{eq: St_quench}
\end{equation}
with the transition probability $P_t(i|j,\theta)= \left| \left< z_i |U_\theta (t) | z_j \right> \right|^2$, initial distribution $P_0(j)=2^{-N}$ and $m_j$ is the total z-magnetization in the eigenstate $\left|z_j \right>$. Consequently, we can extract the spectrum by initializing our system in a product state of z-polarized states after which we quench the system to evolve under $U_\theta(t)$ generated by Hamiltonian $H(\theta)$, and then finally  performing a projective measurement in the z-basis again at time $t$. By repeating the procedure by uniformly sampling initial eigenstates and estimating the product of the initial and final magnetization $m_im_j$, one obtains an estimate of $S(t|\theta)$, see Fig.~\ref{fig:Procedure}. Note that, at this stage, the problem is entirely classical and all quantum physics is hidden in the transition probablity $P_t(i|j,\theta)$. It is the intractability of this transition probability that forms the basis of recent quanum supremacy experiments~\cite{googlesup}. 

In contrast to the latter, we are only interested in estimating a simple statistic, namely the average $m_im_j$. Note that this quantity is bounded by $N^2/4$, hence according to Hoeffding's inequality one needs to sample at most $O(N^4/\epsilon^2)$ times to get an precision of $\epsilon$ on $S(t|\theta)$. At present, the structure of Eq.~\eqref{eq: St_quench} allows one to bound the variance of $m_im_j$ by $3(N/4)^2$, such that $O(N^2/\epsilon^2)$ would suffice. As shown in detail in the SI, one can in general not improve on this scaling with $N$ unless one uses additional structure of the transition probability $P_t$. At short times one for example benefits from importance sampling. While we have no control over the transition probability $P_t$, we can control the initial probability out of which we sample states, as long as those states are easy to prepare.  Since~\eqref{eq: St_quench} is diagonal in the z-basis, it's sufficient to consider sampling product states in z-basis, i.e. one can equivalently write the response function as: 
 \begin{equation}
S(t|\theta)= \sum_{i,j} \left(\frac{m_i m_j P_0(j)}{Q_0(j)} \right) P_t(i|j,\theta)Q_0(j),
\label{eq: St_importance}
\end{equation} 
where $Q_0$ is the distribution from which we sample. By minimizing the variance of estimand $r=m_i m_j P_0/Q_0$, one obtains an optimal sampling distribution. The true optimal depends on time through $P_t$ and since this is unknown to us, we must settle for a good, albeit suboptimal, distribution $Q_0$. Various approximations might be considered, the distribution 
\begin{equation}
Q_0(j)= \frac{4}{N} \frac{m_j^2}{2^{N}},
\end{equation}  
is particularly interesting because it gives zero variance for $r$ the $t=0$ and at any other time the variance is smaller than $(N/4)^2$. Consequently we can estimate $S(t|\theta)$ with precision $\epsilon$ by taking at most $O(N^2/\epsilon^2)$ samples. Given the finite decoherence rate $\gamma$ and the fact that the energy bandwidth of the many-body spectrum scales linearly with $N$, one needs to measure $S(t|\theta)$ at worst in time steps of the order of $1/N$ up to a time that scales as $1/\gamma$. One thus has to repeat the entire circuit at worst $O(N^3/\epsilon^2)$ times. Furthermore, if the time-evolution is implemented as an analog simulation this takes a time of $O(1/\gamma)$. The gate complexity is at worst a factor $N^2$ more because one, at worst, has to implement a Heisenberg interaction between all possible qubits, yielding $O(N^2/\gamma)$. Note that these are worst case scalings, for an extensive spectrum one actually expects linear scaling of the gate complexity with $N$ and typical transition happen between states with only differ by an energy of $O(1)$ such that the typical sampling complexity is only quadratic with $N$.

\section*{Variational Bayesian inference}
\begin{figure}[h]
	\centering
	\includegraphics[width=0.48 \textwidth]{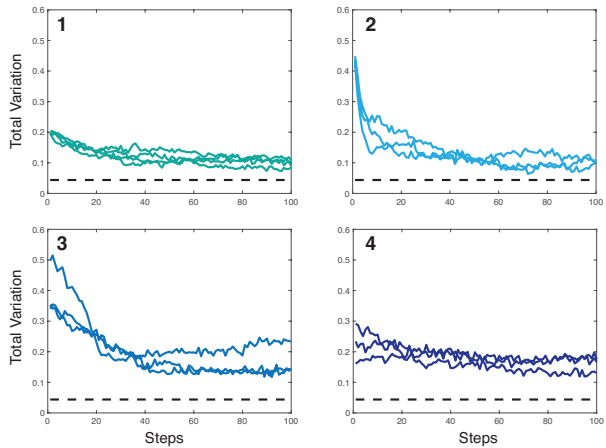}
	\caption{\textbf{Inference} For each of the clusters, labeled according to Fig.~\ref{fig:ClusterSpectra}, we investigate the convergence of the parameter inference in our variational Bayesian inference scheme by looking at the total variation distance between the spectra. The dashed line indicates the shot noise limit, set by the finite number of acquired quantum measurements.}
	\label{fig:Bayes}
\end{figure}

Now that we have a procedure of efficiently obtaining spectra of hypothetical molecules, how do we solve the inference problem? The standard approach would be to do maximum likelihood estimation of the parameters given the experimental spectrum or minimize one of the aforementioned cost functions. This cannot be done analytically and the problem can clearly be highly non-convex. We thus require a method to numerically minimize the error; gradient descend seems an obvious choice but is not well suited for this task.  First of all there is the obvious problem that additional measurements will need to performed to estmate the gradients. Those estimates are not easy to obtain since they require measuring three-point correlators in time. Moreover, using a quantum simulator, one only obtains a statistical estimate of the cost function and its gradient; since we only perform a finite number of measurements. In order to move down the optimization landscape we thus need to resolve the signal from the noise, meaning gradients have to be sufficiently large to be resolved. However, we find extremely small gradients for this problem. Taking for example the Hellinger distance,~$D_H$, used to construct Fig.~\eqref{fig:Distance}, we find the gradient satisfies
\begin{equation}
\left| \partial_\theta D_H^2 \right|=\left| \int \frac{{\rm d\omega}}{2\pi} \sqrt{\frac{\mathcal{A(\omega)}}{A(\omega | \theta)}} \partial_\theta A(\omega |\theta)  \right| \leq \sqrt{I_{\theta \theta}},
\end{equation}
where $I_{\theta \theta}$ is the diagonal component of the Fisher information. The bound simply follows from Cauchy-Schwarz inequality. As shown in the SI, the Fisher information, even for the optimal values, is very small; typically of the order $10^{-4}-10^{-6}$ for our 4 spin molecules. We are thus in a situation of a very shallow rough optimization landscape. The problem is of similar origin as the vanishing gradient problem in quantum neural networks~\cite{mcclean18}.
A gradient free method seems advisable. Here we adopt a Bayesian approach to update our estimated parameters. Alternative approaches, such as the DIRECT method addopted in Ref.~\cite{zoller19}, are expected to work as well. More research on the structure of the optimzation landscape is required to understand the hardness of this inference problem.
Recall Bayes theorem, in the current notation, reads: 
\begin{equation}
P(\theta|\omega)= \frac{A(\omega|\theta)P(\theta)}{A(\omega)},
\end{equation}
where $P(\theta|\omega)$ is the conditional probability to have parameters $\theta$ given that we see spectral weight at frequency $\omega$, $A(\omega|\theta)$ is the NMR spectrum for fixed parameters $\theta$, $P(\theta)$ is the probability to have parameters $\theta$ and $A(\omega)$ is the marginal NMR spectrum averaged over all $\theta$. 
If we acquire some data, say a new spectrum $\mathcal{A}(\omega)$ and we have some prior belief about the distribution $P(\theta)$, we can use it to update our belief about the distribution of the parameters, i.e.
\begin{equation}
P_{i+1}(\theta)= \int \frac{{\rm d}\omega}{2\pi} \mathcal{A}(\omega) \frac{A(\omega|\theta)}{A_i(\omega)}P_i(\theta),
\label{eq:Bayesupdate}
\end{equation}
with $A_i(\omega)=\int {\rm d}\theta A(\omega|\theta)P_i(\theta)$. Note that the above rule indeed conserves positivity and normalization. Moreover, it simply reweights the prior distribution with some weight
\begin{equation}
w_i(\theta)= \int \frac{{\rm d}\omega}{2\pi} \mathcal{A}(\omega) \frac{A(\omega|\theta)}{A_i(\omega)},
\end{equation}
that is directly related to the log-likelihood, since Jensen inequality gives:
\begin{equation}
\log(w_i(\theta)) \geq \int \frac{{\rm d}\omega}{2\pi} \mathcal{A}(\omega)  \log \frac{A(\omega|\theta)}{A_i(\omega)}= \mathcal{L}(\theta) +c,
\end{equation}
where $\mathcal{L(\theta)}$ is the log-likehood and $c$ is a constant independent of $\theta$. Consequently, iterating expression~\eqref{eq:Bayesupdate} is expected to converge to a distribution of parameters which is highly peaked around the maximum likelihood estimate. While it avoids the use of any gradients, it requires us to sample from the current parameter distribution $P_i(\theta)$. This by itself could become intractable and so we make an additional approximation. In order to be able to sample from the parameter distribution, we approximate it by a normal distribution at every step. That is, given that we have obtained some Monte Carlo samples out of $P_i(\theta)$, we can estimate all the weights $w_i(\theta)$ by simply simulating the model and obtaining $A(\omega| \theta_i)$ for all the samples. Next, we approximate $P_{i+1}(\theta)$ with a normal distribution that is a close as possible to it, i.e. has minimal KL-distance. The latter is simply the distribution with the same sample mean and covariance as $P_{i+1}(\theta)$. We use an atomic prior, $P_0(\theta)=\sum_i \frac{1}{N_s} \delta(\theta-\theta_i)$, consisting of all the samples that belong to the same cluster to which the spectrum is identified to belong. The result of this procedure for some randomly chosen test molecules is shown in Fig.~\ref{fig:Bayes}.
We observe steady, albeit noisy, convergence of the molecular spectra. Convergence is limited by three factors, i.e. (i) shot noise from the quantum measurements, (ii) sampling noise from the Monte Carlo procedure and (iii) the Gaussian variational approximation. While both noise sources can be made smaller by using more computational resources, more advanced methods might ultimately be needed. 

\section*{Summary and Outlook}
Here we have presented a method to improve model inference for NMR with relatively modest amount of quantum resources. Similar to generic generative models such as Boltzmann machines, for which a more efficient quantum version has been constructed~\cite{wiebe17,melko18}, we have constructed an application specific model from which a quantum machine can sample more efficiently than a classical computer.  Model parameters are determined through a variational Bayesian approach with an informative prior, constructed by applying t-SNE to a dataset of small molecules.  As a consequence of the noisy nature of the generative model, as well as the absence of significant gradients, both the initial bias as well as the derivative free nature of Bayesian inference are crucial to tackling the problem. This situation, however,  is generic to any hybrid quantum-classical setting that is sufficiently complicated. A similar approach might thus be useful to improve convergence of QAOA or VQE, e.g.heuristic optimization strategies for QAOA have been developed in Ref.~\cite{zhou}.
Both the classical and quantum part of our approach can be extended further. On the quantum side, one can envision developing more efficient approaches for computing the spectra; trading computational time for extra quantum resources. On the classical side, improvements on the inference algorithm might be possible by combining or extending the variational method with Hamiltonian Monte Carlo techniques~\cite{neal}.

It is interesting to extend our technique to other types of experiments. NMR is hardly the only problem where one performs inference on spectroscopic data. For example, one can imagine combining resonant inelastic X-ray scattering (RIXS) data from strongly correlated electron systems~\cite{rixs}, with Fermi-Hubbard simulators based on ultracold atoms~\cite{hofstetter,bloch}. Currently, RIXS data is analyzed by performing numerical studies of small clusters on classical computers (see Ref.~\cite{ament} for review). A DMFT-based hybrid algorithm was recently proposed in~\cite{jaksch}. With cold atoms in optical lattices one may be able to create larger systems and study their non-equilibrium dynamics corresponding to RIXS spectroscopy.

\begin{acknowledgments}
\emph{Acknowledgements.---} DS acknowledges support from the FWO as post-doctoral fellow of the Research Foundation -- Flanders and from a 2019 grant from the Harvard Quantum Initiative Seed Funding program. SM is supported by a research grant from the National Heart, Lung, and Blood Institute (K24 HL136852). OD and HD are supported by a research award from the National Heart, Lung, and Blood Institute, (5K01HL135342) and (T32 HL007575) respectively. ED acknowledges support from the Harvard-MIT CUA, ARO grant number W911NF-20-1-0163, the National Science Foundation through grant No. OAC-1934714, AFOSR Quantum Simulation MURI 
The authors acknowledge useful discussion with P. Mehta, M. Lukin.
\end{acknowledgments}


\bibliography{NMRbib}

\begin{thebibliography}{57}%
\makeatletter
\providecommand \@ifxundefined [1]{%
 \@ifx{#1\undefined}
}%
\providecommand \@ifnum [1]{%
 \ifnum #1\expandafter \@firstoftwo
 \else \expandafter \@secondoftwo
 \fi
}%
\providecommand \@ifx [1]{%
 \ifx #1\expandafter \@firstoftwo
 \else \expandafter \@secondoftwo
 \fi
}%
\providecommand \natexlab [1]{#1}%
\providecommand \enquote  [1]{``#1''}%
\providecommand \bibnamefont  [1]{#1}%
\providecommand \bibfnamefont [1]{#1}%
\providecommand \citenamefont [1]{#1}%
\providecommand \href@noop [0]{\@secondoftwo}%
\providecommand \href [0]{\begingroup \@sanitize@url \@href}%
\providecommand \@href[1]{\@@startlink{#1}\@@href}%
\providecommand \@@href[1]{\endgroup#1\@@endlink}%
\providecommand \@sanitize@url [0]{\catcode `\\12\catcode `\$12\catcode
  `\&12\catcode `\#12\catcode `\^12\catcode `\_12\catcode `\%12\relax}%
\providecommand \@@startlink[1]{}%
\providecommand \@@endlink[0]{}%
\providecommand \url  [0]{\begingroup\@sanitize@url \@url }%
\providecommand \@url [1]{\endgroup\@href {#1}{\urlprefix }}%
\providecommand \urlprefix  [0]{URL }%
\providecommand \Eprint [0]{\href }%
\providecommand \doibase [0]{http://dx.doi.org/}%
\providecommand \selectlanguage [0]{\@gobble}%
\providecommand \bibinfo  [0]{\@secondoftwo}%
\providecommand \bibfield  [0]{\@secondoftwo}%
\providecommand \translation [1]{[#1]}%
\providecommand \BibitemOpen [0]{}%
\providecommand \bibitemStop [0]{}%
\providecommand \bibitemNoStop [0]{.\EOS\space}%
\providecommand \EOS [0]{\spacefactor3000\relax}%
\providecommand \BibitemShut  [1]{\csname bibitem#1\endcsname}%
\let\auto@bib@innerbib\@empty
\bibitem [{\citenamefont {Preskill}(2018)}]{preskill18}%
  \BibitemOpen
  \bibfield  {author} {\bibinfo {author} {\bibfnamefont {J.}~\bibnamefont
  {Preskill}},\ }\href {\doibase 10.22331/q-2018-08-06-79} {\bibfield
  {journal} {\bibinfo  {journal} {{Quantum}}\ }\textbf {\bibinfo {volume}
  {2}},\ \bibinfo {pages} {79} (\bibinfo {year} {2018})}\BibitemShut {NoStop}%
\bibitem [{\citenamefont {Bernien}\ \emph {et~al.}(2017)\citenamefont
  {Bernien}, \citenamefont {Schwartz}, \citenamefont {Keesling}, \citenamefont
  {Levine}, \citenamefont {Omran}, \citenamefont {Pichler}, \citenamefont
  {Choi}, \citenamefont {Zibrov}, \citenamefont {Endres}, \citenamefont
  {Greiner}, \citenamefont {Vuleti{\'c}},\ and\ \citenamefont
  {Lukin}}]{bernien17}%
  \BibitemOpen
  \bibfield  {author} {\bibinfo {author} {\bibfnamefont {H.}~\bibnamefont
  {Bernien}}, \bibinfo {author} {\bibfnamefont {S.}~\bibnamefont {Schwartz}},
  \bibinfo {author} {\bibfnamefont {A.}~\bibnamefont {Keesling}}, \bibinfo
  {author} {\bibfnamefont {H.}~\bibnamefont {Levine}}, \bibinfo {author}
  {\bibfnamefont {A.}~\bibnamefont {Omran}}, \bibinfo {author} {\bibfnamefont
  {H.}~\bibnamefont {Pichler}}, \bibinfo {author} {\bibfnamefont
  {S.}~\bibnamefont {Choi}}, \bibinfo {author} {\bibfnamefont {A.~S.}\
  \bibnamefont {Zibrov}}, \bibinfo {author} {\bibfnamefont {M.}~\bibnamefont
  {Endres}}, \bibinfo {author} {\bibfnamefont {M.}~\bibnamefont {Greiner}},
  \bibinfo {author} {\bibfnamefont {V.}~\bibnamefont {Vuleti{\'c}}}, \ and\
  \bibinfo {author} {\bibfnamefont {M.~D.}\ \bibnamefont {Lukin}},\ }\href
  {https://doi.org/10.1038/nature24622} {\bibfield  {journal} {\bibinfo
  {journal} {Nature}\ }\textbf {\bibinfo {volume} {551}},\ \bibinfo {pages}
  {579 EP } (\bibinfo {year} {2017})}\BibitemShut {NoStop}%
\bibitem [{\citenamefont {Friis}\ \emph {et~al.}(2018)\citenamefont {Friis},
  \citenamefont {Marty}, \citenamefont {Maier}, \citenamefont {Hempel},
  \citenamefont {Holz\"apfel}, \citenamefont {Jurcevic}, \citenamefont
  {Plenio}, \citenamefont {Huber}, \citenamefont {Roos}, \citenamefont
  {Blatt},\ and\ \citenamefont {Lanyon}}]{friis18}%
  \BibitemOpen
  \bibfield  {author} {\bibinfo {author} {\bibfnamefont {N.}~\bibnamefont
  {Friis}}, \bibinfo {author} {\bibfnamefont {O.}~\bibnamefont {Marty}},
  \bibinfo {author} {\bibfnamefont {C.}~\bibnamefont {Maier}}, \bibinfo
  {author} {\bibfnamefont {C.}~\bibnamefont {Hempel}}, \bibinfo {author}
  {\bibfnamefont {M.}~\bibnamefont {Holz\"apfel}}, \bibinfo {author}
  {\bibfnamefont {P.}~\bibnamefont {Jurcevic}}, \bibinfo {author}
  {\bibfnamefont {M.~B.}\ \bibnamefont {Plenio}}, \bibinfo {author}
  {\bibfnamefont {M.}~\bibnamefont {Huber}}, \bibinfo {author} {\bibfnamefont
  {C.}~\bibnamefont {Roos}}, \bibinfo {author} {\bibfnamefont {R.}~\bibnamefont
  {Blatt}}, \ and\ \bibinfo {author} {\bibfnamefont {B.}~\bibnamefont
  {Lanyon}},\ }\href {\doibase 10.1103/PhysRevX.8.021012} {\bibfield  {journal}
  {\bibinfo  {journal} {Phys. Rev. X}\ }\textbf {\bibinfo {volume} {8}},\
  \bibinfo {pages} {021012} (\bibinfo {year} {2018})}\BibitemShut {NoStop}%
\bibitem [{\citenamefont {Farhi}\ \emph {et~al.}(2014)\citenamefont {Farhi},
  \citenamefont {Goldstone},\ and\ \citenamefont {Gutmann}}]{farhi14}%
  \BibitemOpen
  \bibfield  {author} {\bibinfo {author} {\bibfnamefont {E.}~\bibnamefont
  {Farhi}}, \bibinfo {author} {\bibfnamefont {J.}~\bibnamefont {Goldstone}}, \
  and\ \bibinfo {author} {\bibfnamefont {S.}~\bibnamefont {Gutmann}},\
  }\href@noop {} {\enquote {\bibinfo {title} {A quantum approximate
  optimization algorithm},}\ } (\bibinfo {year} {2014}),\ \Eprint
  {http://arxiv.org/abs/arXiv:1411.4028} {arXiv:1411.4028} \BibitemShut
  {NoStop}%
\bibitem [{\citenamefont {Peruzzo}\ \emph {et~al.}(2014)\citenamefont
  {Peruzzo}, \citenamefont {McClean}, \citenamefont {Shadbolt}, \citenamefont
  {Yung}, \citenamefont {Zhou}, \citenamefont {Love}, \citenamefont
  {Aspuru-Guzik},\ and\ \citenamefont {O'Brien}}]{peruzzo14}%
  \BibitemOpen
  \bibfield  {author} {\bibinfo {author} {\bibfnamefont {A.}~\bibnamefont
  {Peruzzo}}, \bibinfo {author} {\bibfnamefont {J.}~\bibnamefont {McClean}},
  \bibinfo {author} {\bibfnamefont {P.}~\bibnamefont {Shadbolt}}, \bibinfo
  {author} {\bibfnamefont {M.-H.}\ \bibnamefont {Yung}}, \bibinfo {author}
  {\bibfnamefont {X.-Q.}\ \bibnamefont {Zhou}}, \bibinfo {author}
  {\bibfnamefont {P.~J.}\ \bibnamefont {Love}}, \bibinfo {author}
  {\bibfnamefont {A.}~\bibnamefont {Aspuru-Guzik}}, \ and\ \bibinfo {author}
  {\bibfnamefont {J.~L.}\ \bibnamefont {O'Brien}},\ }\href
  {https://doi.org/10.1038/ncomms5213} {\bibfield  {journal} {\bibinfo
  {journal} {Nature Communications}\ }\textbf {\bibinfo {volume} {5}},\
  \bibinfo {pages} {4213 EP } (\bibinfo {year} {2014})}\BibitemShut {NoStop}%
\bibitem [{\citenamefont {Kokail}\ \emph
  {et~al.}(2019{\natexlab{a}})\citenamefont {Kokail}, \citenamefont {Maier},
  \citenamefont {van Bijnen}, \citenamefont {Brydges}, \citenamefont {Joshi},
  \citenamefont {Jurcevic}, \citenamefont {Muschik}, \citenamefont {Silvi},
  \citenamefont {Blatt}, \citenamefont {Roos},\ and\ \citenamefont
  {Zoller}}]{kokail19}%
  \BibitemOpen
  \bibfield  {author} {\bibinfo {author} {\bibfnamefont {C.}~\bibnamefont
  {Kokail}}, \bibinfo {author} {\bibfnamefont {C.}~\bibnamefont {Maier}},
  \bibinfo {author} {\bibfnamefont {R.}~\bibnamefont {van Bijnen}}, \bibinfo
  {author} {\bibfnamefont {T.}~\bibnamefont {Brydges}}, \bibinfo {author}
  {\bibfnamefont {M.~K.}\ \bibnamefont {Joshi}}, \bibinfo {author}
  {\bibfnamefont {P.}~\bibnamefont {Jurcevic}}, \bibinfo {author}
  {\bibfnamefont {C.~A.}\ \bibnamefont {Muschik}}, \bibinfo {author}
  {\bibfnamefont {P.}~\bibnamefont {Silvi}}, \bibinfo {author} {\bibfnamefont
  {R.}~\bibnamefont {Blatt}}, \bibinfo {author} {\bibfnamefont {C.~F.}\
  \bibnamefont {Roos}}, \ and\ \bibinfo {author} {\bibfnamefont
  {P.}~\bibnamefont {Zoller}},\ }\href {\doibase 10.1038/s41586-019-1177-4}
  {\bibfield  {journal} {\bibinfo  {journal} {Nature}\ }\textbf {\bibinfo
  {volume} {569}},\ \bibinfo {pages} {355} (\bibinfo {year}
  {2019}{\natexlab{a}})}\BibitemShut {NoStop}%
\bibitem [{\citenamefont {Kandala}\ \emph {et~al.}(2017)\citenamefont
  {Kandala}, \citenamefont {Mezzacapo}, \citenamefont {Temme}, \citenamefont
  {Takita}, \citenamefont {Brink}, \citenamefont {Chow},\ and\ \citenamefont
  {Gambetta}}]{kandala17}%
  \BibitemOpen
  \bibfield  {author} {\bibinfo {author} {\bibfnamefont {A.}~\bibnamefont
  {Kandala}}, \bibinfo {author} {\bibfnamefont {A.}~\bibnamefont {Mezzacapo}},
  \bibinfo {author} {\bibfnamefont {K.}~\bibnamefont {Temme}}, \bibinfo
  {author} {\bibfnamefont {M.}~\bibnamefont {Takita}}, \bibinfo {author}
  {\bibfnamefont {M.}~\bibnamefont {Brink}}, \bibinfo {author} {\bibfnamefont
  {J.~M.}\ \bibnamefont {Chow}}, \ and\ \bibinfo {author} {\bibfnamefont
  {J.~M.}\ \bibnamefont {Gambetta}},\ }\href
  {https://doi.org/10.1038/nature23879} {\bibfield  {journal} {\bibinfo
  {journal} {Nature}\ }\textbf {\bibinfo {volume} {549}},\ \bibinfo {pages}
  {242 EP } (\bibinfo {year} {2017})}\BibitemShut {NoStop}%
\bibitem [{\citenamefont {Colless}\ \emph {et~al.}(2018)\citenamefont
  {Colless}, \citenamefont {Ramasesh}, \citenamefont {Dahlen}, \citenamefont
  {Blok}, \citenamefont {Kimchi-Schwartz}, \citenamefont {McClean},
  \citenamefont {Carter}, \citenamefont {de~Jong},\ and\ \citenamefont
  {Siddiqi}}]{colless18}%
  \BibitemOpen
  \bibfield  {author} {\bibinfo {author} {\bibfnamefont {J.~I.}\ \bibnamefont
  {Colless}}, \bibinfo {author} {\bibfnamefont {V.~V.}\ \bibnamefont
  {Ramasesh}}, \bibinfo {author} {\bibfnamefont {D.}~\bibnamefont {Dahlen}},
  \bibinfo {author} {\bibfnamefont {M.~S.}\ \bibnamefont {Blok}}, \bibinfo
  {author} {\bibfnamefont {M.~E.}\ \bibnamefont {Kimchi-Schwartz}}, \bibinfo
  {author} {\bibfnamefont {J.~R.}\ \bibnamefont {McClean}}, \bibinfo {author}
  {\bibfnamefont {J.}~\bibnamefont {Carter}}, \bibinfo {author} {\bibfnamefont
  {W.~A.}\ \bibnamefont {de~Jong}}, \ and\ \bibinfo {author} {\bibfnamefont
  {I.}~\bibnamefont {Siddiqi}},\ }\href {\doibase 10.1103/PhysRevX.8.011021}
  {\bibfield  {journal} {\bibinfo  {journal} {Phys. Rev. X}\ }\textbf {\bibinfo
  {volume} {8}},\ \bibinfo {pages} {011021} (\bibinfo {year}
  {2018})}\BibitemShut {NoStop}%
\bibitem [{\citenamefont {Diggle}\ and\ \citenamefont
  {Gratton}(1984)}]{diggle84}%
  \BibitemOpen
  \bibfield  {author} {\bibinfo {author} {\bibfnamefont {P.~J.}\ \bibnamefont
  {Diggle}}\ and\ \bibinfo {author} {\bibfnamefont {R.~J.}\ \bibnamefont
  {Gratton}},\ }\href@noop {} {\bibfield  {journal} {\bibinfo  {journal}
  {Journal of the Royal Statistical Society, Series B: Methodological}\
  }\textbf {\bibinfo {volume} {46}},\ \bibinfo {pages} {193} (\bibinfo {year}
  {1984})}\BibitemShut {NoStop}%
\bibitem [{\citenamefont {Beaumont}\ \emph {et~al.}(2002)\citenamefont
  {Beaumont}, \citenamefont {Zhang},\ and\ \citenamefont
  {Balding}}]{beaumont02}%
  \BibitemOpen
  \bibfield  {author} {\bibinfo {author} {\bibfnamefont {M.~A.}\ \bibnamefont
  {Beaumont}}, \bibinfo {author} {\bibfnamefont {W.}~\bibnamefont {Zhang}}, \
  and\ \bibinfo {author} {\bibfnamefont {D.~J.}\ \bibnamefont {Balding}},\
  }\href {https://www.genetics.org/content/162/4/2025} {\bibfield  {journal}
  {\bibinfo  {journal} {Genetics}\ }\textbf {\bibinfo {volume} {162}},\
  \bibinfo {pages} {2025} (\bibinfo {year} {2002})},\ \Eprint
  {http://arxiv.org/abs/https://www.genetics.org/content/162/4/2025.full.pdf}
  {https://www.genetics.org/content/162/4/2025.full.pdf} \BibitemShut {NoStop}%
\bibitem [{\citenamefont {Gershenfeld}\ and\ \citenamefont
  {Chuang}(1997)}]{gershenfeld350}%
  \BibitemOpen
  \bibfield  {author} {\bibinfo {author} {\bibfnamefont {N.~A.}\ \bibnamefont
  {Gershenfeld}}\ and\ \bibinfo {author} {\bibfnamefont {I.~L.}\ \bibnamefont
  {Chuang}},\ }\href {\doibase 10.1126/science.275.5298.350} {\bibfield
  {journal} {\bibinfo  {journal} {Science}\ }\textbf {\bibinfo {volume}
  {275}},\ \bibinfo {pages} {350} (\bibinfo {year} {1997})}\BibitemShut
  {NoStop}%
\bibitem [{\citenamefont {Braunstein}\ \emph {et~al.}(1999)\citenamefont
  {Braunstein}, \citenamefont {Caves}, \citenamefont {Jozsa}, \citenamefont
  {Linden}, \citenamefont {Popescu},\ and\ \citenamefont
  {Schack}}]{braunstein99}%
  \BibitemOpen
  \bibfield  {author} {\bibinfo {author} {\bibfnamefont {S.~L.}\ \bibnamefont
  {Braunstein}}, \bibinfo {author} {\bibfnamefont {C.~M.}\ \bibnamefont
  {Caves}}, \bibinfo {author} {\bibfnamefont {R.}~\bibnamefont {Jozsa}},
  \bibinfo {author} {\bibfnamefont {N.}~\bibnamefont {Linden}}, \bibinfo
  {author} {\bibfnamefont {S.}~\bibnamefont {Popescu}}, \ and\ \bibinfo
  {author} {\bibfnamefont {R.}~\bibnamefont {Schack}},\ }\href {\doibase
  10.1103/PhysRevLett.83.1054} {\bibfield  {journal} {\bibinfo  {journal}
  {Phys. Rev. Lett.}\ }\textbf {\bibinfo {volume} {83}},\ \bibinfo {pages}
  {1054} (\bibinfo {year} {1999})}\BibitemShut {NoStop}%
\bibitem [{\citenamefont {Menicucci}\ and\ \citenamefont
  {Caves}(2002)}]{menicucci02}%
  \BibitemOpen
  \bibfield  {author} {\bibinfo {author} {\bibfnamefont {N.~C.}\ \bibnamefont
  {Menicucci}}\ and\ \bibinfo {author} {\bibfnamefont {C.~M.}\ \bibnamefont
  {Caves}},\ }\href {\doibase 10.1103/PhysRevLett.88.167901} {\bibfield
  {journal} {\bibinfo  {journal} {Phys. Rev. Lett.}\ }\textbf {\bibinfo
  {volume} {88}},\ \bibinfo {pages} {167901} (\bibinfo {year}
  {2002})}\BibitemShut {NoStop}%
\bibitem [{\citenamefont {Datta}\ and\ \citenamefont {Vidal}(2007)}]{datta07}%
  \BibitemOpen
  \bibfield  {author} {\bibinfo {author} {\bibfnamefont {A.}~\bibnamefont
  {Datta}}\ and\ \bibinfo {author} {\bibfnamefont {G.}~\bibnamefont {Vidal}},\
  }\href {\doibase 10.1103/PhysRevA.75.042310} {\bibfield  {journal} {\bibinfo
  {journal} {Phys. Rev. A}\ }\textbf {\bibinfo {volume} {75}},\ \bibinfo
  {pages} {042310} (\bibinfo {year} {2007})}\BibitemShut {NoStop}%
\bibitem [{\citenamefont {Knill}\ and\ \citenamefont
  {Laflamme}(1998)}]{knill98}%
  \BibitemOpen
  \bibfield  {author} {\bibinfo {author} {\bibfnamefont {E.}~\bibnamefont
  {Knill}}\ and\ \bibinfo {author} {\bibfnamefont {R.}~\bibnamefont
  {Laflamme}},\ }\href {\doibase 10.1103/PhysRevLett.81.5672} {\bibfield
  {journal} {\bibinfo  {journal} {Phys. Rev. Lett.}\ }\textbf {\bibinfo
  {volume} {81}},\ \bibinfo {pages} {5672} (\bibinfo {year}
  {1998})}\BibitemShut {NoStop}%
\bibitem [{\citenamefont {Biamonte}\ \emph {et~al.}(2017)\citenamefont
  {Biamonte}, \citenamefont {Wittek}, \citenamefont {Pancotti}, \citenamefont
  {Rebentrost}, \citenamefont {Wiebe},\ and\ \citenamefont
  {Lloyd}}]{biamonte17}%
  \BibitemOpen
  \bibfield  {author} {\bibinfo {author} {\bibfnamefont {J.}~\bibnamefont
  {Biamonte}}, \bibinfo {author} {\bibfnamefont {P.}~\bibnamefont {Wittek}},
  \bibinfo {author} {\bibfnamefont {N.}~\bibnamefont {Pancotti}}, \bibinfo
  {author} {\bibfnamefont {P.}~\bibnamefont {Rebentrost}}, \bibinfo {author}
  {\bibfnamefont {N.}~\bibnamefont {Wiebe}}, \ and\ \bibinfo {author}
  {\bibfnamefont {S.}~\bibnamefont {Lloyd}},\ }\href
  {https://doi.org/10.1038/nature23474} {\bibfield  {journal} {\bibinfo
  {journal} {Nature}\ }\textbf {\bibinfo {volume} {549}},\ \bibinfo {pages}
  {195 EP } (\bibinfo {year} {2017})}\BibitemShut {NoStop}%
\bibitem [{\citenamefont {{Brassard}}\ and\ \citenamefont
  {{Hoyer}}(1997)}]{brassard}%
  \BibitemOpen
  \bibfield  {author} {\bibinfo {author} {\bibfnamefont {G.}~\bibnamefont
  {{Brassard}}}\ and\ \bibinfo {author} {\bibfnamefont {P.}~\bibnamefont
  {{Hoyer}}},\ }in\ \href {\doibase 10.1109/ISTCS.1997.595153} {\emph {\bibinfo
  {booktitle} {Proceedings of the Fifth Israeli Symposium on Theory of
  Computing and Systems}}}\ (\bibinfo {year} {1997})\ pp.\ \bibinfo {pages}
  {12--23}\BibitemShut {NoStop}%
\bibitem [{\citenamefont {Grover}(1998)}]{grover}%
  \BibitemOpen
  \bibfield  {author} {\bibinfo {author} {\bibfnamefont {L.~K.}\ \bibnamefont
  {Grover}},\ }\href {\doibase 10.1103/PhysRevLett.80.4329} {\bibfield
  {journal} {\bibinfo  {journal} {Phys. Rev. Lett.}\ }\textbf {\bibinfo
  {volume} {80}},\ \bibinfo {pages} {4329} (\bibinfo {year}
  {1998})}\BibitemShut {NoStop}%
\bibitem [{\citenamefont {Harrow}\ \emph {et~al.}(2009)\citenamefont {Harrow},
  \citenamefont {Hassidim},\ and\ \citenamefont {Lloyd}}]{hhl}%
  \BibitemOpen
  \bibfield  {author} {\bibinfo {author} {\bibfnamefont {A.~W.}\ \bibnamefont
  {Harrow}}, \bibinfo {author} {\bibfnamefont {A.}~\bibnamefont {Hassidim}}, \
  and\ \bibinfo {author} {\bibfnamefont {S.}~\bibnamefont {Lloyd}},\ }\href
  {\doibase 10.1103/PhysRevLett.103.150502} {\bibfield  {journal} {\bibinfo
  {journal} {Phys. Rev. Lett.}\ }\textbf {\bibinfo {volume} {103}},\ \bibinfo
  {pages} {150502} (\bibinfo {year} {2009})}\BibitemShut {NoStop}%
\bibitem [{\citenamefont {Bothwell}\ and\ \citenamefont
  {Griffin}(2011)}]{bothwell11}%
  \BibitemOpen
  \bibfield  {author} {\bibinfo {author} {\bibfnamefont {J.~H.~F.}\
  \bibnamefont {Bothwell}}\ and\ \bibinfo {author} {\bibfnamefont {J.~L.}\
  \bibnamefont {Griffin}},\ }\href {\doibase 10.1111/j.1469-185X.2010.00157.x}
  {\bibfield  {journal} {\bibinfo  {journal} {Biological Reviews}\ }\textbf
  {\bibinfo {volume} {86}},\ \bibinfo {pages} {493} (\bibinfo {year} {2011})},\
  \Eprint
  {http://arxiv.org/abs/https://onlinelibrary.wiley.com/doi/pdf/10.1111/j.1469-185X.2010.00157.x}
  {https://onlinelibrary.wiley.com/doi/pdf/10.1111/j.1469-185X.2010.00157.x}
  \BibitemShut {NoStop}%
\bibitem [{\citenamefont {Hwang}\ and\ \citenamefont {Choi}(2015)}]{jonghee19}%
  \BibitemOpen
  \bibfield  {author} {\bibinfo {author} {\bibfnamefont {J.-H.}\ \bibnamefont
  {Hwang}}\ and\ \bibinfo {author} {\bibfnamefont {C.~S.}\ \bibnamefont
  {Choi}},\ }\href {https://doi.org/10.1038/emm.2014.101} {\bibfield  {journal}
  {\bibinfo  {journal} {Experimental \&Amp; Molecular Medicine}\ }\textbf
  {\bibinfo {volume} {47}},\ \bibinfo {pages} {e139 EP } (\bibinfo {year}
  {2015})}\BibitemShut {NoStop}%
\bibitem [{\citenamefont {Beckonert}\ \emph {et~al.}(2007)\citenamefont
  {Beckonert}, \citenamefont {Keun}, \citenamefont {Ebbels}, \citenamefont
  {Bundy}, \citenamefont {Holmes}, \citenamefont {Lindon},\ and\ \citenamefont
  {Nicholson}}]{olaf07}%
  \BibitemOpen
  \bibfield  {author} {\bibinfo {author} {\bibfnamefont {O.}~\bibnamefont
  {Beckonert}}, \bibinfo {author} {\bibfnamefont {H.~C.}\ \bibnamefont {Keun}},
  \bibinfo {author} {\bibfnamefont {T.~M.~D.}\ \bibnamefont {Ebbels}}, \bibinfo
  {author} {\bibfnamefont {J.}~\bibnamefont {Bundy}}, \bibinfo {author}
  {\bibfnamefont {E.}~\bibnamefont {Holmes}}, \bibinfo {author} {\bibfnamefont
  {J.~C.}\ \bibnamefont {Lindon}}, \ and\ \bibinfo {author} {\bibfnamefont
  {J.~K.}\ \bibnamefont {Nicholson}},\ }\href
  {https://doi.org/10.1038/nprot.2007.376} {\bibfield  {journal} {\bibinfo
  {journal} {Nature Protocols}\ }\textbf {\bibinfo {volume} {2}},\ \bibinfo
  {pages} {2692 EP } (\bibinfo {year} {2007})}\BibitemShut {NoStop}%
\bibitem [{\citenamefont {Larive}\ \emph {et~al.}(2015)\citenamefont {Larive},
  \citenamefont {Barding},\ and\ \citenamefont {Dinges}}]{larive15}%
  \BibitemOpen
  \bibfield  {author} {\bibinfo {author} {\bibfnamefont {C.~K.}\ \bibnamefont
  {Larive}}, \bibinfo {author} {\bibfnamefont {G.~A.}\ \bibnamefont {Barding}},
  \ and\ \bibinfo {author} {\bibfnamefont {M.~M.}\ \bibnamefont {Dinges}},\
  }\bibfield  {booktitle} {\emph {\bibinfo {booktitle} {Analytical
  Chemistry}},\ }\href {\doibase 10.1021/ac504075g} {\bibfield  {journal}
  {\bibinfo  {journal} {Analytical Chemistry}\ }\textbf {\bibinfo {volume}
  {87}},\ \bibinfo {pages} {133} (\bibinfo {year} {2015})}\BibitemShut
  {NoStop}%
\bibitem [{\citenamefont {Napolitano}\ \emph {et~al.}(2013)\citenamefont
  {Napolitano}, \citenamefont {Lankin}, \citenamefont {McAlpine}, \citenamefont
  {Niemitz}, \citenamefont {Korhonen}, \citenamefont {Chen},\ and\
  \citenamefont {Pauli}}]{napolitano13}%
  \BibitemOpen
  \bibfield  {author} {\bibinfo {author} {\bibfnamefont {J.}~\bibnamefont
  {Napolitano}}, \bibinfo {author} {\bibfnamefont {D.~C.}\ \bibnamefont
  {Lankin}}, \bibinfo {author} {\bibfnamefont {J.~B.}\ \bibnamefont
  {McAlpine}}, \bibinfo {author} {\bibfnamefont {M.}~\bibnamefont {Niemitz}},
  \bibinfo {author} {\bibfnamefont {S.-P.}\ \bibnamefont {Korhonen}}, \bibinfo
  {author} {\bibfnamefont {S.-N.}\ \bibnamefont {Chen}}, \ and\ \bibinfo
  {author} {\bibfnamefont {G.~F.}\ \bibnamefont {Pauli}},\ }\bibfield
  {booktitle} {\emph {\bibinfo {booktitle} {The Journal of Organic
  Chemistry}},\ }\href {\doibase 10.1021/jo4011624} {\bibfield  {journal}
  {\bibinfo  {journal} {The Journal of Organic Chemistry}\ }\textbf {\bibinfo
  {volume} {78}},\ \bibinfo {pages} {9963} (\bibinfo {year}
  {2013})}\BibitemShut {NoStop}%
\bibitem [{\citenamefont {Ravanbakhsh}\ \emph {et~al.}(2015)\citenamefont
  {Ravanbakhsh}, \citenamefont {Liu}, \citenamefont {Bjordahl}, \citenamefont
  {Mandal}, \citenamefont {Grant}, \citenamefont {Wilson}, \citenamefont
  {Eisner}, \citenamefont {Sinelnikov}, \citenamefont {Hu}, \citenamefont
  {Luchinat}, \citenamefont {Greiner},\ and\ \citenamefont
  {Wishart}}]{ravanbakhsh15}%
  \BibitemOpen
  \bibfield  {author} {\bibinfo {author} {\bibfnamefont {S.}~\bibnamefont
  {Ravanbakhsh}}, \bibinfo {author} {\bibfnamefont {P.}~\bibnamefont {Liu}},
  \bibinfo {author} {\bibfnamefont {T.~C.}\ \bibnamefont {Bjordahl}}, \bibinfo
  {author} {\bibfnamefont {R.}~\bibnamefont {Mandal}}, \bibinfo {author}
  {\bibfnamefont {J.~R.}\ \bibnamefont {Grant}}, \bibinfo {author}
  {\bibfnamefont {M.}~\bibnamefont {Wilson}}, \bibinfo {author} {\bibfnamefont
  {R.}~\bibnamefont {Eisner}}, \bibinfo {author} {\bibfnamefont
  {I.}~\bibnamefont {Sinelnikov}}, \bibinfo {author} {\bibfnamefont
  {X.}~\bibnamefont {Hu}}, \bibinfo {author} {\bibfnamefont {C.}~\bibnamefont
  {Luchinat}}, \bibinfo {author} {\bibfnamefont {R.}~\bibnamefont {Greiner}}, \
  and\ \bibinfo {author} {\bibfnamefont {D.~S.}\ \bibnamefont {Wishart}},\
  }\href {\doibase 10.1371/journal.pone.0124219} {\bibfield  {journal}
  {\bibinfo  {journal} {PLOS ONE}\ }\textbf {\bibinfo {volume} {10}},\ \bibinfo
  {pages} {1} (\bibinfo {year} {2015})}\BibitemShut {NoStop}%
\bibitem [{\citenamefont {De~Graaf}\ and\ \citenamefont
  {Bove\'e}(1990)}]{degraaf90}%
  \BibitemOpen
  \bibfield  {author} {\bibinfo {author} {\bibfnamefont {A.~A.}\ \bibnamefont
  {De~Graaf}}\ and\ \bibinfo {author} {\bibfnamefont {W.~M. M.~J.}\
  \bibnamefont {Bove\'e}},\ }\href {\doibase 10.1002/mrm.1910150212} {\bibfield
   {journal} {\bibinfo  {journal} {Magnetic Resonance in Medicine}\ }\textbf
  {\bibinfo {volume} {15}},\ \bibinfo {pages} {305} (\bibinfo {year}
  {1990})}\BibitemShut {NoStop}%
\bibitem [{\citenamefont {Wevers}\ \emph {et~al.}(1994)\citenamefont {Wevers},
  \citenamefont {Engelke},\ and\ \citenamefont {Heerschap}}]{wevers94}%
  \BibitemOpen
  \bibfield  {author} {\bibinfo {author} {\bibfnamefont {R.~A.}\ \bibnamefont
  {Wevers}}, \bibinfo {author} {\bibfnamefont {U.}~\bibnamefont {Engelke}}, \
  and\ \bibinfo {author} {\bibfnamefont {A.}~\bibnamefont {Heerschap}},\ }\href
  {http://clinchem.aaccjnls.org/content/40/7/1245} {\bibfield  {journal}
  {\bibinfo  {journal} {Clinical Chemistry}\ }\textbf {\bibinfo {volume}
  {40}},\ \bibinfo {pages} {1245} (\bibinfo {year} {1994})},\ \Eprint
  {http://arxiv.org/abs/http://clinchem.aaccjnls.org/content/40/7/1245.full.pdf}
  {http://clinchem.aaccjnls.org/content/40/7/1245.full.pdf} \BibitemShut
  {NoStop}%
\bibitem [{\citenamefont {Wevers}\ \emph {et~al.}(1995)\citenamefont {Wevers},
  \citenamefont {Engelke}, \citenamefont {Wendel}, \citenamefont {de~Jong},
  \citenamefont {Gabre{\"e}ls},\ and\ \citenamefont {Heerschap}}]{wevers95}%
  \BibitemOpen
  \bibfield  {author} {\bibinfo {author} {\bibfnamefont {R.~A.}\ \bibnamefont
  {Wevers}}, \bibinfo {author} {\bibfnamefont {U.}~\bibnamefont {Engelke}},
  \bibinfo {author} {\bibfnamefont {U.}~\bibnamefont {Wendel}}, \bibinfo
  {author} {\bibfnamefont {J.~G.}\ \bibnamefont {de~Jong}}, \bibinfo {author}
  {\bibfnamefont {F.~J.}\ \bibnamefont {Gabre{\"e}ls}}, \ and\ \bibinfo
  {author} {\bibfnamefont {A.}~\bibnamefont {Heerschap}},\ }\href
  {http://clinchem.aaccjnls.org/content/41/5/744} {\bibfield  {journal}
  {\bibinfo  {journal} {Clinical Chemistry}\ }\textbf {\bibinfo {volume}
  {41}},\ \bibinfo {pages} {744} (\bibinfo {year} {1995})},\ \Eprint
  {http://arxiv.org/abs/http://clinchem.aaccjnls.org/content/41/5/744.full.pdf}
  {http://clinchem.aaccjnls.org/content/41/5/744.full.pdf} \BibitemShut
  {NoStop}%
\bibitem [{\citenamefont {Govindaraju}\ \emph {et~al.}(2000)\citenamefont
  {Govindaraju}, \citenamefont {Young},\ and\ \citenamefont
  {Maudsley}}]{govindaraju00}%
  \BibitemOpen
  \bibfield  {author} {\bibinfo {author} {\bibfnamefont {V.}~\bibnamefont
  {Govindaraju}}, \bibinfo {author} {\bibfnamefont {K.}~\bibnamefont {Young}},
  \ and\ \bibinfo {author} {\bibfnamefont {A.~A.}\ \bibnamefont {Maudsley}},\
  }\href {\doibase 10.1002/1099-1492(200005)13:3<129::AID-NBM619>3.0.CO;2-V}
  {\bibfield  {journal} {\bibinfo  {journal} {NMR in Biomedicine}\ }\textbf
  {\bibinfo {volume} {13}},\ \bibinfo {pages} {129} (\bibinfo {year}
  {2000})}\BibitemShut {NoStop}%
\bibitem [{\citenamefont {Dashti}\ \emph {et~al.}(2018)\citenamefont {Dashti},
  \citenamefont {Wedell}, \citenamefont {Westler}, \citenamefont {Tonelli},
  \citenamefont {Aceti}, \citenamefont {Amarasinghe}, \citenamefont {Markley},\
  and\ \citenamefont {Eghbalnia}}]{hesam1}%
  \BibitemOpen
  \bibfield  {author} {\bibinfo {author} {\bibfnamefont {H.}~\bibnamefont
  {Dashti}}, \bibinfo {author} {\bibfnamefont {J.~R.}\ \bibnamefont {Wedell}},
  \bibinfo {author} {\bibfnamefont {W.~M.}\ \bibnamefont {Westler}}, \bibinfo
  {author} {\bibfnamefont {M.}~\bibnamefont {Tonelli}}, \bibinfo {author}
  {\bibfnamefont {D.}~\bibnamefont {Aceti}}, \bibinfo {author} {\bibfnamefont
  {G.~K.}\ \bibnamefont {Amarasinghe}}, \bibinfo {author} {\bibfnamefont
  {J.~L.}\ \bibnamefont {Markley}}, \ and\ \bibinfo {author} {\bibfnamefont
  {H.~R.}\ \bibnamefont {Eghbalnia}},\ }\bibfield  {booktitle} {\emph {\bibinfo
  {booktitle} {Analytical Chemistry}},\ }\href {\doibase
  10.1021/acs.analchem.8b02660} {\bibfield  {journal} {\bibinfo  {journal}
  {Analytical Chemistry}\ }\textbf {\bibinfo {volume} {90}},\ \bibinfo {pages}
  {10646} (\bibinfo {year} {2018})}\BibitemShut {NoStop}%
\bibitem [{\citenamefont {Dashti}\ \emph {et~al.}(2017)\citenamefont {Dashti},
  \citenamefont {Westler}, \citenamefont {Tonelli}, \citenamefont {Wedell},
  \citenamefont {Markley},\ and\ \citenamefont {Eghbalnia}}]{hesam2}%
  \BibitemOpen
  \bibfield  {author} {\bibinfo {author} {\bibfnamefont {H.}~\bibnamefont
  {Dashti}}, \bibinfo {author} {\bibfnamefont {W.~M.}\ \bibnamefont {Westler}},
  \bibinfo {author} {\bibfnamefont {M.}~\bibnamefont {Tonelli}}, \bibinfo
  {author} {\bibfnamefont {J.~R.}\ \bibnamefont {Wedell}}, \bibinfo {author}
  {\bibfnamefont {J.~L.}\ \bibnamefont {Markley}}, \ and\ \bibinfo {author}
  {\bibfnamefont {H.~R.}\ \bibnamefont {Eghbalnia}},\ }\bibfield  {booktitle}
  {\emph {\bibinfo {booktitle} {Analytical Chemistry}},\ }\href {\doibase
  10.1021/acs.analchem.7b02884} {\bibfield  {journal} {\bibinfo  {journal}
  {Analytical Chemistry}\ }\textbf {\bibinfo {volume} {89}},\ \bibinfo {pages}
  {12201} (\bibinfo {year} {2017})}\BibitemShut {NoStop}%
\bibitem [{\citenamefont {Pickard}\ and\ \citenamefont
  {Mauri}(2001)}]{chemshift}%
  \BibitemOpen
  \bibfield  {author} {\bibinfo {author} {\bibfnamefont {C.~J.}\ \bibnamefont
  {Pickard}}\ and\ \bibinfo {author} {\bibfnamefont {F.}~\bibnamefont
  {Mauri}},\ }\href {\doibase 10.1103/PhysRevB.63.245101} {\bibfield  {journal}
  {\bibinfo  {journal} {Phys. Rev. B}\ }\textbf {\bibinfo {volume} {63}},\
  \bibinfo {pages} {245101} (\bibinfo {year} {2001})}\BibitemShut {NoStop}%
\bibitem [{\citenamefont {Paruzzo}\ \emph {et~al.}(2018)\citenamefont
  {Paruzzo}, \citenamefont {Hofstetter}, \citenamefont {Musil}, \citenamefont
  {De}, \citenamefont {Ceriotti},\ and\ \citenamefont {Emsley}}]{chemshiftAI}%
  \BibitemOpen
  \bibfield  {author} {\bibinfo {author} {\bibfnamefont {F.~M.}\ \bibnamefont
  {Paruzzo}}, \bibinfo {author} {\bibfnamefont {A.}~\bibnamefont {Hofstetter}},
  \bibinfo {author} {\bibfnamefont {F.}~\bibnamefont {Musil}}, \bibinfo
  {author} {\bibfnamefont {S.}~\bibnamefont {De}}, \bibinfo {author}
  {\bibfnamefont {M.}~\bibnamefont {Ceriotti}}, \ and\ \bibinfo {author}
  {\bibfnamefont {L.}~\bibnamefont {Emsley}},\ }\href {\doibase
  10.1038/s41467-018-06972-x} {\bibfield  {journal} {\bibinfo  {journal}
  {Nature Communications}\ }\textbf {\bibinfo {volume} {9}},\ \bibinfo {pages}
  {4501} (\bibinfo {year} {2018})}\BibitemShut {NoStop}%
\bibitem [{\citenamefont {Levitt}(2008)}]{levitt}%
  \BibitemOpen
  \bibfield  {author} {\bibinfo {author} {\bibfnamefont {M.~H.}\ \bibnamefont
  {Levitt}},\ }\href@noop {} {\emph {\bibinfo {title} {Spin Dynamics: Basics of
  Nuclear Magnetic Resonance}}}\ (\bibinfo  {publisher} {Wiley},\ \bibinfo
  {year} {2008})\BibitemShut {NoStop}%
\bibitem [{\citenamefont {Dashti}(2019)}]{hesam3}%
  \BibitemOpen
  \bibfield  {author} {\bibinfo {author} {\bibfnamefont {H.}~\bibnamefont
  {Dashti}},\ }\href {http://gissmo.nmrfam.wisc.edu/} {\enquote {\bibinfo
  {title} {http://gissmo.nmrfam.wisc.edu/},}\ } (\bibinfo {year}
  {2019})\BibitemShut {NoStop}%
\bibitem [{\citenamefont {van~der Maaten}\ and\ \citenamefont
  {Hinton}(2008)}]{tsne08}%
  \BibitemOpen
  \bibfield  {author} {\bibinfo {author} {\bibfnamefont {L.}~\bibnamefont
  {van~der Maaten}}\ and\ \bibinfo {author} {\bibfnamefont {G.}~\bibnamefont
  {Hinton}},\ }\href {http://www.jmlr.org/papers/v9/vandermaaten08a.html}
  {\bibfield  {journal} {\bibinfo  {journal} {Journal of Machine Learning
  Research}\ }\textbf {\bibinfo {volume} {9}},\ \bibinfo {pages} {2579}
  (\bibinfo {year} {2008})}\BibitemShut {NoStop}%
\bibitem [{\citenamefont {van~der Maaten}(2019)}]{lvdm19}%
  \BibitemOpen
  \bibfield  {author} {\bibinfo {author} {\bibfnamefont {L.}~\bibnamefont
  {van~der Maaten}},\ }\href {https://lvdmaaten.github.io/tsne/} {\enquote
  {\bibinfo {title} {https://lvdmaaten.github.io/tsne/},}\ } (\bibinfo {year}
  {2019})\BibitemShut {NoStop}%
\bibitem [{\citenamefont {Ester}\ \emph {et~al.}(1996)\citenamefont {Ester},
  \citenamefont {Kriegel}, \citenamefont {Sander},\ and\ \citenamefont
  {Xu}}]{dbscan}%
  \BibitemOpen
  \bibfield  {author} {\bibinfo {author} {\bibfnamefont {M.}~\bibnamefont
  {Ester}}, \bibinfo {author} {\bibfnamefont {H.-P.}\ \bibnamefont {Kriegel}},
  \bibinfo {author} {\bibfnamefont {J.}~\bibnamefont {Sander}}, \ and\ \bibinfo
  {author} {\bibfnamefont {X.}~\bibnamefont {Xu}}\ }(\bibinfo  {publisher}
  {AAAI Press},\ \bibinfo {year} {1996})\ pp.\ \bibinfo {pages}
  {226--231}\BibitemShut {NoStop}%
\bibitem [{\citenamefont {Ulrich}\ \emph {et~al.}(2008)\citenamefont {Ulrich},
  \citenamefont {Akutsu}, \citenamefont {Doreleijers}, \citenamefont {Harano},
  \citenamefont {Ioannidis}, \citenamefont {Lin}, \citenamefont {Livny},
  \citenamefont {Mading}, \citenamefont {Maziuk}, \citenamefont {Miller},
  \citenamefont {Nakatani}, \citenamefont {Schulte}, \citenamefont {Tolmie},
  \citenamefont {Kent~Wenger}, \citenamefont {Yao},\ and\ \citenamefont
  {Markley}}]{ulrich08}%
  \BibitemOpen
  \bibfield  {author} {\bibinfo {author} {\bibfnamefont {E.~L.}\ \bibnamefont
  {Ulrich}}, \bibinfo {author} {\bibfnamefont {H.}~\bibnamefont {Akutsu}},
  \bibinfo {author} {\bibfnamefont {J.~F.}\ \bibnamefont {Doreleijers}},
  \bibinfo {author} {\bibfnamefont {Y.}~\bibnamefont {Harano}}, \bibinfo
  {author} {\bibfnamefont {Y.~E.}\ \bibnamefont {Ioannidis}}, \bibinfo {author}
  {\bibfnamefont {J.}~\bibnamefont {Lin}}, \bibinfo {author} {\bibfnamefont
  {M.}~\bibnamefont {Livny}}, \bibinfo {author} {\bibfnamefont
  {S.}~\bibnamefont {Mading}}, \bibinfo {author} {\bibfnamefont
  {D.}~\bibnamefont {Maziuk}}, \bibinfo {author} {\bibfnamefont
  {Z.}~\bibnamefont {Miller}}, \bibinfo {author} {\bibfnamefont
  {E.}~\bibnamefont {Nakatani}}, \bibinfo {author} {\bibfnamefont {C.~F.}\
  \bibnamefont {Schulte}}, \bibinfo {author} {\bibfnamefont {D.~E.}\
  \bibnamefont {Tolmie}}, \bibinfo {author} {\bibfnamefont {R.}~\bibnamefont
  {Kent~Wenger}}, \bibinfo {author} {\bibfnamefont {H.}~\bibnamefont {Yao}}, \
  and\ \bibinfo {author} {\bibfnamefont {J.~L.}\ \bibnamefont {Markley}},\
  }\href {\doibase 10.1093/nar/gkm957} {\bibfield  {journal} {\bibinfo
  {journal} {Nucleic acids research}\ }\textbf {\bibinfo {volume} {36}},\
  \bibinfo {pages} {D402} (\bibinfo {year} {2008})}\BibitemShut {NoStop}%
\bibitem [{\citenamefont {Sokolenko}\ \emph {et~al.}(2019)\citenamefont
  {Sokolenko}, \citenamefont {JÃ©zÃ©quel}, \citenamefont {Hajjar},
  \citenamefont {Farjon}, \citenamefont {Akoka},\ and\ \citenamefont
  {Giraudeau}}]{sokolenko19}%
  \BibitemOpen
  \bibfield  {author} {\bibinfo {author} {\bibfnamefont {S.}~\bibnamefont
  {Sokolenko}}, \bibinfo {author} {\bibfnamefont {T.}~\bibnamefont
  {JÃ©zÃ©quel}}, \bibinfo {author} {\bibfnamefont {G.}~\bibnamefont
  {Hajjar}}, \bibinfo {author} {\bibfnamefont {J.}~\bibnamefont {Farjon}},
  \bibinfo {author} {\bibfnamefont {S.}~\bibnamefont {Akoka}}, \ and\ \bibinfo
  {author} {\bibfnamefont {P.}~\bibnamefont {Giraudeau}},\ }\href {\doibase
  https://doi.org/10.1016/j.jmr.2018.11.004} {\bibfield  {journal} {\bibinfo
  {journal} {Journal of Magnetic Resonance}\ }\textbf {\bibinfo {volume}
  {298}},\ \bibinfo {pages} {91 } (\bibinfo {year} {2019})}\BibitemShut
  {NoStop}%
\bibitem [{\citenamefont {{Xu}}\ \emph {et~al.}(2019)\citenamefont {{Xu}},
  \citenamefont {{Marrelec}}, \citenamefont {{Bernard}},\ and\ \citenamefont
  {{Grimal}}}]{ku19}%
  \BibitemOpen
  \bibfield  {author} {\bibinfo {author} {\bibfnamefont {K.}~\bibnamefont
  {{Xu}}}, \bibinfo {author} {\bibfnamefont {G.}~\bibnamefont {{Marrelec}}},
  \bibinfo {author} {\bibfnamefont {S.}~\bibnamefont {{Bernard}}}, \ and\
  \bibinfo {author} {\bibfnamefont {Q.}~\bibnamefont {{Grimal}}},\ }\href
  {\doibase 10.1109/TSP.2018.2878543} {\bibfield  {journal} {\bibinfo
  {journal} {IEEE Transactions on Signal Processing}\ }\textbf {\bibinfo
  {volume} {67}},\ \bibinfo {pages} {4} (\bibinfo {year} {2019})}\BibitemShut
  {NoStop}%
\bibitem [{\citenamefont {Arute}\ \emph {et~al.}(2019)\citenamefont {Arute},
  \citenamefont {Arya}, \citenamefont {Babbush}, \citenamefont {Bacon},
  \citenamefont {Bardin}, \citenamefont {Barends}, \citenamefont {Biswas},
  \citenamefont {Boixo}, \citenamefont {Brandao}, \citenamefont {Buell},
  \citenamefont {Burkett}, \citenamefont {Chen}, \citenamefont {Chen},
  \citenamefont {Chiaro}, \citenamefont {Collins}, \citenamefont {Courtney},
  \citenamefont {Dunsworth}, \citenamefont {Farhi}, \citenamefont {Foxen},
  \citenamefont {Fowler}, \citenamefont {Gidney}, \citenamefont {Giustina},
  \citenamefont {Graff}, \citenamefont {Guerin}, \citenamefont {Habegger},
  \citenamefont {Harrigan}, \citenamefont {Hartmann}, \citenamefont {Ho},
  \citenamefont {Hoffmann}, \citenamefont {Huang}, \citenamefont {Humble},
  \citenamefont {Isakov}, \citenamefont {Jeffrey}, \citenamefont {Jiang},
  \citenamefont {Kafri}, \citenamefont {Kechedzhi}, \citenamefont {Kelly},
  \citenamefont {Klimov}, \citenamefont {Knysh}, \citenamefont {Korotkov},
  \citenamefont {Kostritsa}, \citenamefont {Landhuis}, \citenamefont
  {Lindmark}, \citenamefont {Lucero}, \citenamefont {Lyakh}, \citenamefont
  {Mandr{\`a}}, \citenamefont {McClean}, \citenamefont {McEwen}, \citenamefont
  {Megrant}, \citenamefont {Mi}, \citenamefont {Michielsen}, \citenamefont
  {Mohseni}, \citenamefont {Mutus}, \citenamefont {Naaman}, \citenamefont
  {Neeley}, \citenamefont {Neill}, \citenamefont {Niu}, \citenamefont {Ostby},
  \citenamefont {Petukhov}, \citenamefont {Platt}, \citenamefont {Quintana},
  \citenamefont {Rieffel}, \citenamefont {Roushan}, \citenamefont {Rubin},
  \citenamefont {Sank}, \citenamefont {Satzinger}, \citenamefont {Smelyanskiy},
  \citenamefont {Sung}, \citenamefont {Trevithick}, \citenamefont
  {Vainsencher}, \citenamefont {Villalonga}, \citenamefont {White},
  \citenamefont {Yao}, \citenamefont {Yeh}, \citenamefont {Zalcman},
  \citenamefont {Neven},\ and\ \citenamefont {Martinis}}]{googlesup}%
  \BibitemOpen
  \bibfield  {author} {\bibinfo {author} {\bibfnamefont {F.}~\bibnamefont
  {Arute}}, \bibinfo {author} {\bibfnamefont {K.}~\bibnamefont {Arya}},
  \bibinfo {author} {\bibfnamefont {R.}~\bibnamefont {Babbush}}, \bibinfo
  {author} {\bibfnamefont {D.}~\bibnamefont {Bacon}}, \bibinfo {author}
  {\bibfnamefont {J.~C.}\ \bibnamefont {Bardin}}, \bibinfo {author}
  {\bibfnamefont {R.}~\bibnamefont {Barends}}, \bibinfo {author} {\bibfnamefont
  {R.}~\bibnamefont {Biswas}}, \bibinfo {author} {\bibfnamefont
  {S.}~\bibnamefont {Boixo}}, \bibinfo {author} {\bibfnamefont {F.~G. S.~L.}\
  \bibnamefont {Brandao}}, \bibinfo {author} {\bibfnamefont {D.~A.}\
  \bibnamefont {Buell}}, \bibinfo {author} {\bibfnamefont {B.}~\bibnamefont
  {Burkett}}, \bibinfo {author} {\bibfnamefont {Y.}~\bibnamefont {Chen}},
  \bibinfo {author} {\bibfnamefont {Z.}~\bibnamefont {Chen}}, \bibinfo {author}
  {\bibfnamefont {B.}~\bibnamefont {Chiaro}}, \bibinfo {author} {\bibfnamefont
  {R.}~\bibnamefont {Collins}}, \bibinfo {author} {\bibfnamefont
  {W.}~\bibnamefont {Courtney}}, \bibinfo {author} {\bibfnamefont
  {A.}~\bibnamefont {Dunsworth}}, \bibinfo {author} {\bibfnamefont
  {E.}~\bibnamefont {Farhi}}, \bibinfo {author} {\bibfnamefont
  {B.}~\bibnamefont {Foxen}}, \bibinfo {author} {\bibfnamefont
  {A.}~\bibnamefont {Fowler}}, \bibinfo {author} {\bibfnamefont
  {C.}~\bibnamefont {Gidney}}, \bibinfo {author} {\bibfnamefont
  {M.}~\bibnamefont {Giustina}}, \bibinfo {author} {\bibfnamefont
  {R.}~\bibnamefont {Graff}}, \bibinfo {author} {\bibfnamefont
  {K.}~\bibnamefont {Guerin}}, \bibinfo {author} {\bibfnamefont
  {S.}~\bibnamefont {Habegger}}, \bibinfo {author} {\bibfnamefont {M.~P.}\
  \bibnamefont {Harrigan}}, \bibinfo {author} {\bibfnamefont {M.~J.}\
  \bibnamefont {Hartmann}}, \bibinfo {author} {\bibfnamefont {A.}~\bibnamefont
  {Ho}}, \bibinfo {author} {\bibfnamefont {M.}~\bibnamefont {Hoffmann}},
  \bibinfo {author} {\bibfnamefont {T.}~\bibnamefont {Huang}}, \bibinfo
  {author} {\bibfnamefont {T.~S.}\ \bibnamefont {Humble}}, \bibinfo {author}
  {\bibfnamefont {S.~V.}\ \bibnamefont {Isakov}}, \bibinfo {author}
  {\bibfnamefont {E.}~\bibnamefont {Jeffrey}}, \bibinfo {author} {\bibfnamefont
  {Z.}~\bibnamefont {Jiang}}, \bibinfo {author} {\bibfnamefont
  {D.}~\bibnamefont {Kafri}}, \bibinfo {author} {\bibfnamefont
  {K.}~\bibnamefont {Kechedzhi}}, \bibinfo {author} {\bibfnamefont
  {J.}~\bibnamefont {Kelly}}, \bibinfo {author} {\bibfnamefont {P.~V.}\
  \bibnamefont {Klimov}}, \bibinfo {author} {\bibfnamefont {S.}~\bibnamefont
  {Knysh}}, \bibinfo {author} {\bibfnamefont {A.}~\bibnamefont {Korotkov}},
  \bibinfo {author} {\bibfnamefont {F.}~\bibnamefont {Kostritsa}}, \bibinfo
  {author} {\bibfnamefont {D.}~\bibnamefont {Landhuis}}, \bibinfo {author}
  {\bibfnamefont {M.}~\bibnamefont {Lindmark}}, \bibinfo {author}
  {\bibfnamefont {E.}~\bibnamefont {Lucero}}, \bibinfo {author} {\bibfnamefont
  {D.}~\bibnamefont {Lyakh}}, \bibinfo {author} {\bibfnamefont
  {S.}~\bibnamefont {Mandr{\`a}}}, \bibinfo {author} {\bibfnamefont {J.~R.}\
  \bibnamefont {McClean}}, \bibinfo {author} {\bibfnamefont {M.}~\bibnamefont
  {McEwen}}, \bibinfo {author} {\bibfnamefont {A.}~\bibnamefont {Megrant}},
  \bibinfo {author} {\bibfnamefont {X.}~\bibnamefont {Mi}}, \bibinfo {author}
  {\bibfnamefont {K.}~\bibnamefont {Michielsen}}, \bibinfo {author}
  {\bibfnamefont {M.}~\bibnamefont {Mohseni}}, \bibinfo {author} {\bibfnamefont
  {J.}~\bibnamefont {Mutus}}, \bibinfo {author} {\bibfnamefont
  {O.}~\bibnamefont {Naaman}}, \bibinfo {author} {\bibfnamefont
  {M.}~\bibnamefont {Neeley}}, \bibinfo {author} {\bibfnamefont
  {C.}~\bibnamefont {Neill}}, \bibinfo {author} {\bibfnamefont {M.~Y.}\
  \bibnamefont {Niu}}, \bibinfo {author} {\bibfnamefont {E.}~\bibnamefont
  {Ostby}}, \bibinfo {author} {\bibfnamefont {A.}~\bibnamefont {Petukhov}},
  \bibinfo {author} {\bibfnamefont {J.~C.}\ \bibnamefont {Platt}}, \bibinfo
  {author} {\bibfnamefont {C.}~\bibnamefont {Quintana}}, \bibinfo {author}
  {\bibfnamefont {E.~G.}\ \bibnamefont {Rieffel}}, \bibinfo {author}
  {\bibfnamefont {P.}~\bibnamefont {Roushan}}, \bibinfo {author} {\bibfnamefont
  {N.~C.}\ \bibnamefont {Rubin}}, \bibinfo {author} {\bibfnamefont
  {D.}~\bibnamefont {Sank}}, \bibinfo {author} {\bibfnamefont {K.~J.}\
  \bibnamefont {Satzinger}}, \bibinfo {author} {\bibfnamefont {V.}~\bibnamefont
  {Smelyanskiy}}, \bibinfo {author} {\bibfnamefont {K.~J.}\ \bibnamefont
  {Sung}}, \bibinfo {author} {\bibfnamefont {M.~D.}\ \bibnamefont
  {Trevithick}}, \bibinfo {author} {\bibfnamefont {A.}~\bibnamefont
  {Vainsencher}}, \bibinfo {author} {\bibfnamefont {B.}~\bibnamefont
  {Villalonga}}, \bibinfo {author} {\bibfnamefont {T.}~\bibnamefont {White}},
  \bibinfo {author} {\bibfnamefont {Z.~J.}\ \bibnamefont {Yao}}, \bibinfo
  {author} {\bibfnamefont {P.}~\bibnamefont {Yeh}}, \bibinfo {author}
  {\bibfnamefont {A.}~\bibnamefont {Zalcman}}, \bibinfo {author} {\bibfnamefont
  {H.}~\bibnamefont {Neven}}, \ and\ \bibinfo {author} {\bibfnamefont {J.~M.}\
  \bibnamefont {Martinis}},\ }\href {\doibase 10.1038/s41586-019-1666-5}
  {\bibfield  {journal} {\bibinfo  {journal} {Nature}\ }\textbf {\bibinfo
  {volume} {574}},\ \bibinfo {pages} {505} (\bibinfo {year}
  {2019})}\BibitemShut {NoStop}%
\bibitem [{\citenamefont {McClean}\ \emph {et~al.}(2018)\citenamefont
  {McClean}, \citenamefont {Boixo}, \citenamefont {Smelyanskiy}, \citenamefont
  {Babbush},\ and\ \citenamefont {Neven}}]{mcclean18}%
  \BibitemOpen
  \bibfield  {author} {\bibinfo {author} {\bibfnamefont {J.~R.}\ \bibnamefont
  {McClean}}, \bibinfo {author} {\bibfnamefont {S.}~\bibnamefont {Boixo}},
  \bibinfo {author} {\bibfnamefont {V.~N.}\ \bibnamefont {Smelyanskiy}},
  \bibinfo {author} {\bibfnamefont {R.}~\bibnamefont {Babbush}}, \ and\
  \bibinfo {author} {\bibfnamefont {H.}~\bibnamefont {Neven}},\ }\href
  {\doibase 10.1038/s41467-018-07090-4} {\bibfield  {journal} {\bibinfo
  {journal} {Nature Communications}\ }\textbf {\bibinfo {volume} {9}},\
  \bibinfo {pages} {4812} (\bibinfo {year} {2018})}\BibitemShut {NoStop}%
\bibitem [{\citenamefont {Kokail}\ \emph
  {et~al.}(2019{\natexlab{b}})\citenamefont {Kokail}, \citenamefont {Maier},
  \citenamefont {van Bijnen}, \citenamefont {Brydges}, \citenamefont {Joshi},
  \citenamefont {Jurcevic}, \citenamefont {Muschik}, \citenamefont {Silvi},
  \citenamefont {Blatt}, \citenamefont {Roos},\ and\ \citenamefont
  {Zoller}}]{zoller19}%
  \BibitemOpen
  \bibfield  {author} {\bibinfo {author} {\bibfnamefont {C.}~\bibnamefont
  {Kokail}}, \bibinfo {author} {\bibfnamefont {C.}~\bibnamefont {Maier}},
  \bibinfo {author} {\bibfnamefont {R.}~\bibnamefont {van Bijnen}}, \bibinfo
  {author} {\bibfnamefont {T.}~\bibnamefont {Brydges}}, \bibinfo {author}
  {\bibfnamefont {M.~K.}\ \bibnamefont {Joshi}}, \bibinfo {author}
  {\bibfnamefont {P.}~\bibnamefont {Jurcevic}}, \bibinfo {author}
  {\bibfnamefont {C.~A.}\ \bibnamefont {Muschik}}, \bibinfo {author}
  {\bibfnamefont {P.}~\bibnamefont {Silvi}}, \bibinfo {author} {\bibfnamefont
  {R.}~\bibnamefont {Blatt}}, \bibinfo {author} {\bibfnamefont {C.~F.}\
  \bibnamefont {Roos}}, \ and\ \bibinfo {author} {\bibfnamefont
  {P.}~\bibnamefont {Zoller}},\ }\href {\doibase 10.1038/s41586-019-1177-4}
  {\bibfield  {journal} {\bibinfo  {journal} {Nature}\ }\textbf {\bibinfo
  {volume} {569}},\ \bibinfo {pages} {355} (\bibinfo {year}
  {2019}{\natexlab{b}})}\BibitemShut {NoStop}%
\bibitem [{\citenamefont {Kieferov\'a}\ and\ \citenamefont
  {Wiebe}(2017)}]{wiebe17}%
  \BibitemOpen
  \bibfield  {author} {\bibinfo {author} {\bibfnamefont {M.}~\bibnamefont
  {Kieferov\'a}}\ and\ \bibinfo {author} {\bibfnamefont {N.}~\bibnamefont
  {Wiebe}},\ }\href {\doibase 10.1103/PhysRevA.96.062327} {\bibfield  {journal}
  {\bibinfo  {journal} {Phys. Rev. A}\ }\textbf {\bibinfo {volume} {96}},\
  \bibinfo {pages} {062327} (\bibinfo {year} {2017})}\BibitemShut {NoStop}%
\bibitem [{\citenamefont {Amin}\ \emph {et~al.}(2018)\citenamefont {Amin},
  \citenamefont {Andriyash}, \citenamefont {Rolfe}, \citenamefont
  {Kulchytskyy},\ and\ \citenamefont {Melko}}]{melko18}%
  \BibitemOpen
  \bibfield  {author} {\bibinfo {author} {\bibfnamefont {M.~H.}\ \bibnamefont
  {Amin}}, \bibinfo {author} {\bibfnamefont {E.}~\bibnamefont {Andriyash}},
  \bibinfo {author} {\bibfnamefont {J.}~\bibnamefont {Rolfe}}, \bibinfo
  {author} {\bibfnamefont {B.}~\bibnamefont {Kulchytskyy}}, \ and\ \bibinfo
  {author} {\bibfnamefont {R.}~\bibnamefont {Melko}},\ }\href {\doibase
  10.1103/PhysRevX.8.021050} {\bibfield  {journal} {\bibinfo  {journal} {Phys.
  Rev. X}\ }\textbf {\bibinfo {volume} {8}},\ \bibinfo {pages} {021050}
  (\bibinfo {year} {2018})}\BibitemShut {NoStop}%
\bibitem [{\citenamefont {Zhou}\ \emph {et~al.}(2018)\citenamefont {Zhou},
  \citenamefont {Wang}, \citenamefont {Choi}, \citenamefont {Pichler},\ and\
  \citenamefont {Lukin}}]{zhou}%
  \BibitemOpen
  \bibfield  {author} {\bibinfo {author} {\bibfnamefont {L.}~\bibnamefont
  {Zhou}}, \bibinfo {author} {\bibfnamefont {S.-T.}\ \bibnamefont {Wang}},
  \bibinfo {author} {\bibfnamefont {S.}~\bibnamefont {Choi}}, \bibinfo {author}
  {\bibfnamefont {H.}~\bibnamefont {Pichler}}, \ and\ \bibinfo {author}
  {\bibfnamefont {M.~D.}\ \bibnamefont {Lukin}},\ }\href@noop {} {\enquote
  {\bibinfo {title} {Quantum approximate optimization algorithm: Performance,
  mechanism, and implementation on near-term devices},}\ } (\bibinfo {year}
  {2018}),\ \Eprint {http://arxiv.org/abs/arXiv:1812.01041} {arXiv:1812.01041}
  \BibitemShut {NoStop}%
\bibitem [{\citenamefont {Neal}(2012)}]{neal}%
  \BibitemOpen
  \bibfield  {author} {\bibinfo {author} {\bibfnamefont {R.~M.}\ \bibnamefont
  {Neal}},\ }\href@noop {} {\  (\bibinfo {year} {2012})},\ \Eprint
  {http://arxiv.org/abs/arXiv:1206.1901} {arXiv:1206.1901} \BibitemShut
  {NoStop}%
\bibitem [{\citenamefont {Murakami}\ and\ \citenamefont
  {Ishihara}(2017)}]{rixs}%
  \BibitemOpen
  \bibinfo {editor} {\bibfnamefont {Y.}~\bibnamefont {Murakami}}\ and\ \bibinfo
  {editor} {\bibfnamefont {S.}~\bibnamefont {Ishihara}},\ eds.,\ \href
  {\doibase 10.1007/978-3-662-53227-0} {\emph {\bibinfo {title} {Resonant X-Ray
  Scattering in Correlated Systems}}}\ (\bibinfo  {publisher} {Springer Berlin
  Heidelberg},\ \bibinfo {year} {2017})\BibitemShut {NoStop}%
\bibitem [{\citenamefont {Hofstetter}\ \emph {et~al.}(2002)\citenamefont
  {Hofstetter}, \citenamefont {Cirac}, \citenamefont {Zoller}, \citenamefont
  {Demler},\ and\ \citenamefont {Lukin}}]{hofstetter}%
  \BibitemOpen
  \bibfield  {author} {\bibinfo {author} {\bibfnamefont {W.}~\bibnamefont
  {Hofstetter}}, \bibinfo {author} {\bibfnamefont {J.~I.}\ \bibnamefont
  {Cirac}}, \bibinfo {author} {\bibfnamefont {P.}~\bibnamefont {Zoller}},
  \bibinfo {author} {\bibfnamefont {E.}~\bibnamefont {Demler}}, \ and\ \bibinfo
  {author} {\bibfnamefont {M.~D.}\ \bibnamefont {Lukin}},\ }\href {\doibase
  10.1103/PhysRevLett.89.220407} {\bibfield  {journal} {\bibinfo  {journal}
  {Phys. Rev. Lett.}\ }\textbf {\bibinfo {volume} {89}},\ \bibinfo {pages}
  {220407} (\bibinfo {year} {2002})}\BibitemShut {NoStop}%
\bibitem [{\citenamefont {Bloch}\ \emph {et~al.}(2008)\citenamefont {Bloch},
  \citenamefont {Dalibard},\ and\ \citenamefont {Zwerger}}]{bloch}%
  \BibitemOpen
  \bibfield  {author} {\bibinfo {author} {\bibfnamefont {I.}~\bibnamefont
  {Bloch}}, \bibinfo {author} {\bibfnamefont {J.}~\bibnamefont {Dalibard}}, \
  and\ \bibinfo {author} {\bibfnamefont {W.}~\bibnamefont {Zwerger}},\ }\href
  {\doibase 10.1103/RevModPhys.80.885} {\bibfield  {journal} {\bibinfo
  {journal} {Rev. Mod. Phys.}\ }\textbf {\bibinfo {volume} {80}},\ \bibinfo
  {pages} {885} (\bibinfo {year} {2008})}\BibitemShut {NoStop}%
\bibitem [{\citenamefont {Ament}\ \emph {et~al.}(2011)\citenamefont {Ament},
  \citenamefont {van Veenendaal}, \citenamefont {Devereaux}, \citenamefont
  {Hill},\ and\ \citenamefont {van~den Brink}}]{ament}%
  \BibitemOpen
  \bibfield  {author} {\bibinfo {author} {\bibfnamefont {L.~J.~P.}\
  \bibnamefont {Ament}}, \bibinfo {author} {\bibfnamefont {M.}~\bibnamefont
  {van Veenendaal}}, \bibinfo {author} {\bibfnamefont {T.~P.}\ \bibnamefont
  {Devereaux}}, \bibinfo {author} {\bibfnamefont {J.~P.}\ \bibnamefont {Hill}},
  \ and\ \bibinfo {author} {\bibfnamefont {J.}~\bibnamefont {van~den Brink}},\
  }\href {\doibase 10.1103/RevModPhys.83.705} {\bibfield  {journal} {\bibinfo
  {journal} {Rev. Mod. Phys.}\ }\textbf {\bibinfo {volume} {83}},\ \bibinfo
  {pages} {705} (\bibinfo {year} {2011})}\BibitemShut {NoStop}%
\bibitem [{\citenamefont {Kreula}\ \emph {et~al.}(2016)\citenamefont {Kreula},
  \citenamefont {Garc{\'\i}a-{\'A}lvarez}, \citenamefont {Lamata},
  \citenamefont {Clark}, \citenamefont {Solano},\ and\ \citenamefont
  {Jaksch}}]{jaksch}%
  \BibitemOpen
  \bibfield  {author} {\bibinfo {author} {\bibfnamefont {J.~M.}\ \bibnamefont
  {Kreula}}, \bibinfo {author} {\bibfnamefont {L.}~\bibnamefont
  {Garc{\'\i}a-{\'A}lvarez}}, \bibinfo {author} {\bibfnamefont
  {L.}~\bibnamefont {Lamata}}, \bibinfo {author} {\bibfnamefont {S.~R.}\
  \bibnamefont {Clark}}, \bibinfo {author} {\bibfnamefont {E.}~\bibnamefont
  {Solano}}, \ and\ \bibinfo {author} {\bibfnamefont {D.}~\bibnamefont
  {Jaksch}},\ }\href {\doibase 10.1140/epjqt/s40507-016-0049-1} {\bibfield
  {journal} {\bibinfo  {journal} {EPJ Quantum Technology}\ }\textbf {\bibinfo
  {volume} {3}},\ \bibinfo {pages} {11} (\bibinfo {year} {2016})}\BibitemShut
  {NoStop}%
\bibitem [{\citenamefont {Mehta}\ \emph {et~al.}(2019)\citenamefont {Mehta},
  \citenamefont {Bukov}, \citenamefont {Wang}, \citenamefont {Day},
  \citenamefont {Richardson}, \citenamefont {Fisher},\ and\ \citenamefont
  {Schwab}}]{pankaj19}%
  \BibitemOpen
  \bibfield  {author} {\bibinfo {author} {\bibfnamefont {P.}~\bibnamefont
  {Mehta}}, \bibinfo {author} {\bibfnamefont {M.}~\bibnamefont {Bukov}},
  \bibinfo {author} {\bibfnamefont {C.-H.}\ \bibnamefont {Wang}}, \bibinfo
  {author} {\bibfnamefont {A.~G.}\ \bibnamefont {Day}}, \bibinfo {author}
  {\bibfnamefont {C.}~\bibnamefont {Richardson}}, \bibinfo {author}
  {\bibfnamefont {C.~K.}\ \bibnamefont {Fisher}}, \ and\ \bibinfo {author}
  {\bibfnamefont {D.~J.}\ \bibnamefont {Schwab}},\ }\href {\doibase
  https://doi.org/10.1016/j.physrep.2019.03.001} {\bibfield  {journal}
  {\bibinfo  {journal} {Physics Reports}\ }\textbf {\bibinfo {volume} {810}},\
  \bibinfo {pages} {1 } (\bibinfo {year} {2019})},\ \bibinfo {note} {a
  high-bias, low-variance introduction to Machine Learning for
  physicists}\BibitemShut {NoStop}%
\bibitem [{\citenamefont {Jeffreys}(1946)}]{jeffreys46}%
  \BibitemOpen
  \bibfield  {author} {\bibinfo {author} {\bibfnamefont {H.}~\bibnamefont
  {Jeffreys}},\ }\href {\doibase 10.1098/rspa.1946.0056} {\bibfield  {journal}
  {\bibinfo  {journal} {Proceedings of the Royal Society of London. Series A.
  Mathematical and Physical Sciences}\ }\textbf {\bibinfo {volume} {186}},\
  \bibinfo {pages} {453} (\bibinfo {year} {1946})}\BibitemShut {NoStop}%
\bibitem [{\citenamefont {Waterfall}\ \emph {et~al.}(2006)\citenamefont
  {Waterfall}, \citenamefont {Casey}, \citenamefont {Gutenkunst}, \citenamefont
  {Brown}, \citenamefont {Myers}, \citenamefont {Brouwer}, \citenamefont
  {Elser},\ and\ \citenamefont {Sethna}}]{waterfall06}%
  \BibitemOpen
  \bibfield  {author} {\bibinfo {author} {\bibfnamefont {J.~J.}\ \bibnamefont
  {Waterfall}}, \bibinfo {author} {\bibfnamefont {F.~P.}\ \bibnamefont
  {Casey}}, \bibinfo {author} {\bibfnamefont {R.~N.}\ \bibnamefont
  {Gutenkunst}}, \bibinfo {author} {\bibfnamefont {K.~S.}\ \bibnamefont
  {Brown}}, \bibinfo {author} {\bibfnamefont {C.~R.}\ \bibnamefont {Myers}},
  \bibinfo {author} {\bibfnamefont {P.~W.}\ \bibnamefont {Brouwer}}, \bibinfo
  {author} {\bibfnamefont {V.}~\bibnamefont {Elser}}, \ and\ \bibinfo {author}
  {\bibfnamefont {J.~P.}\ \bibnamefont {Sethna}},\ }\href {\doibase
  10.1103/PhysRevLett.97.150601} {\bibfield  {journal} {\bibinfo  {journal}
  {Phys. Rev. Lett.}\ }\textbf {\bibinfo {volume} {97}},\ \bibinfo {pages}
  {150601} (\bibinfo {year} {2006})}\BibitemShut {NoStop}%
\bibitem [{\citenamefont {Machta}\ \emph {et~al.}(2013)\citenamefont {Machta},
  \citenamefont {Chachra}, \citenamefont {Transtrum},\ and\ \citenamefont
  {Sethna}}]{machta13}%
  \BibitemOpen
  \bibfield  {author} {\bibinfo {author} {\bibfnamefont {B.~B.}\ \bibnamefont
  {Machta}}, \bibinfo {author} {\bibfnamefont {R.}~\bibnamefont {Chachra}},
  \bibinfo {author} {\bibfnamefont {M.~K.}\ \bibnamefont {Transtrum}}, \ and\
  \bibinfo {author} {\bibfnamefont {J.~P.}\ \bibnamefont {Sethna}},\ }\href
  {\doibase 10.1126/science.1238723} {\bibfield  {journal} {\bibinfo  {journal}
  {Science}\ }\textbf {\bibinfo {volume} {342}},\ \bibinfo {pages} {604}
  (\bibinfo {year} {2013})}\BibitemShut {NoStop}%
\end{thebibliography}%

\clearpage

\setcounter{page}{1}
\renewcommand{\thepage}{S\arabic{page}}

\setcounter{table}{0}
\renewcommand{\thetable}{S\arabic{table}}
\renewcommand{\theHtable}{Supplement.\thetable}

\setcounter{figure}{0}
\renewcommand\thefigure{S\arabic{figure}}
\renewcommand{\theHfigure}{Supplement.\thefigure}

\setcounter{equation}{0}
\renewcommand{\theequation}{S\arabic{equation}}
\renewcommand{\theHequation}{Supplement.\theequation}

\setcounter{section}{0}
\renewcommand{\thesection}{S\arabic{section}}
\renewcommand{\theHsection}{Supplement.\thesection}


\onecolumngrid
\begin{center}
	\noindent\textbf{Supplemental Material for:}
	\bigskip
		
	\noindent\textbf{\large{Quantum approximate Bayesian computation for NMR model inference}}
		
\end{center}

\section{\label{sec:simal} Simularity measures}
\begin{figure}[h]
	\centering
	\includegraphics[width=0.9 \textwidth]{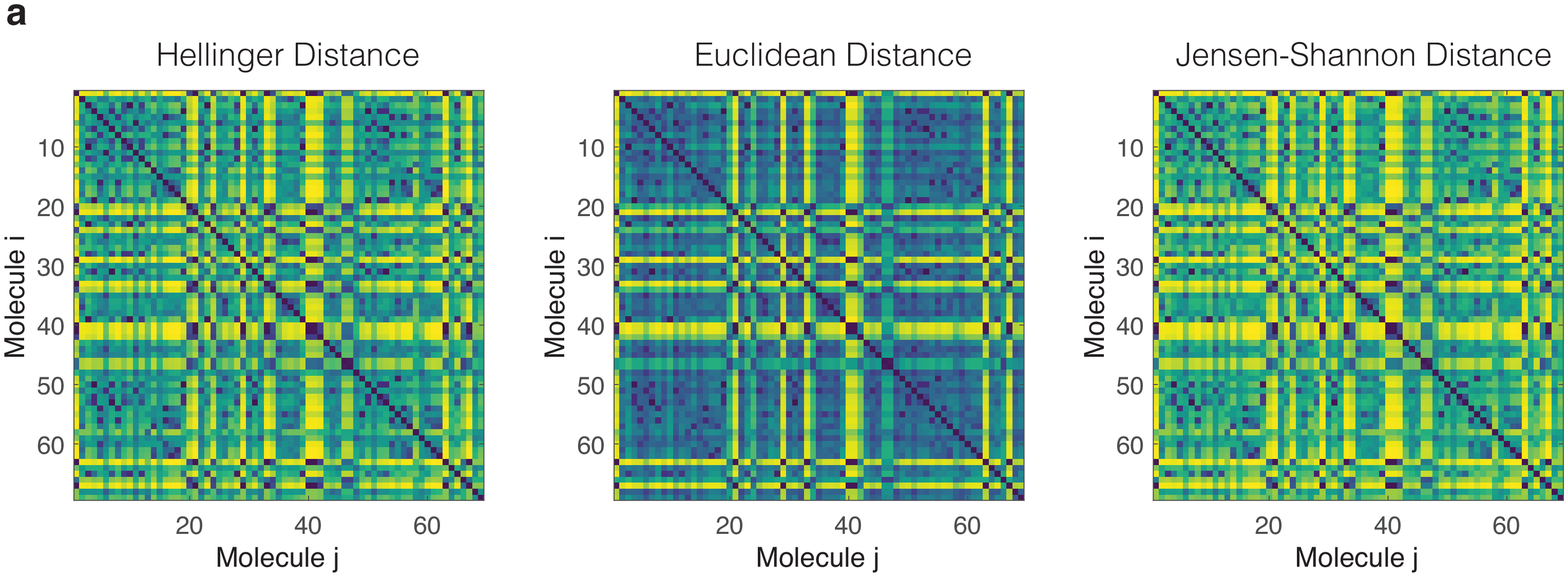}
	\includegraphics[width=0.9 \textwidth]{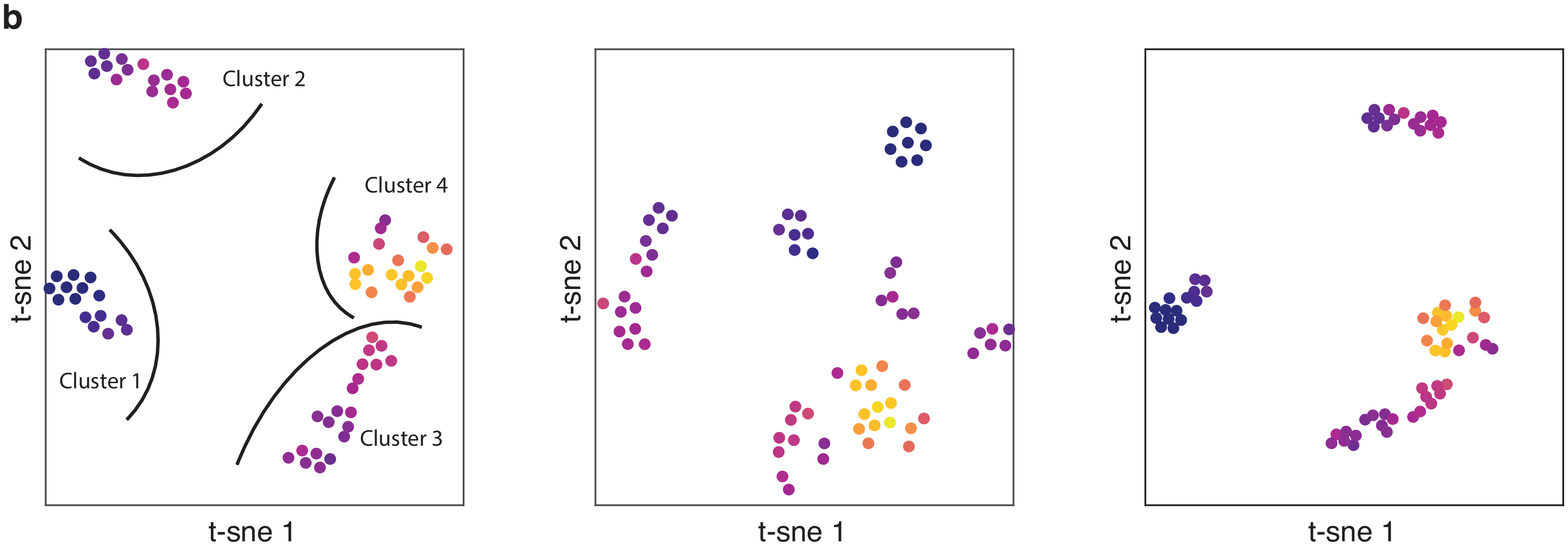}
	\caption{ \textbf{Clustering} In order to indenfity whether naturally occuring molecules have some atypical NMR spectrum we perform a clustering analysis based on 3 different measures of similarity. In panel--A we show the distance between the various NMR spectra for three different distant metrices.  To extract clusters we perform a t-SNE shown in panel--B for each of the metrices respectively. The t-SNE is performed with the same initial seed and perplexity(10) for all plots. The KL-loss for the shown plots was $\{0.145,0.510,0.299\}$ for the Hellinger, Euclidean and JS distance respectively. }
	\label{fig:DistanceSI}
\end{figure}

To perform clustering or simply find the best fit to a certain spectrum one has to define a measure of distance or equivalently of simularity between different spectra. A priori, there is no unique optimal chooice for this and certain measures might be much better suited for the current problem then others. Let's therefore have a closer look at a few possible distance matrices:
\begin{eqnarray}
{\rm Euclidean:} \, D^2_2(i,j) &=& \int \frac{{\rm d}\omega}{2\pi} \left(A_i-A_j\right)^2, \nonumber \\
{\rm Hellinger:} \, D^2_H(i,j)&=&\frac{1}{2} \int  \frac{{\rm d}\omega}{2\pi} \left(\sqrt{A_i}-\sqrt{A_j}\right)^2, \nonumber \\
{\rm Jensen-Shannon:} \, D^2_{JS}(i,j)&=&\frac{1}{2} \int \frac{{\rm d}\omega}{2\pi} \left(A_i \log A_i +A_j \log A_j- (A_i+A_j)\log \frac{A_i+A_j}{2} \right), \nonumber
\end{eqnarray}
where $A_i$ is short hand notation denoting $A_i=A(\omega|\theta_i)$. Note that the spectrum is positive and can be normalized since it satisfies the f-sum rule:
\begin{equation}
S(0|\theta)= \int \frac{{\rm d}\omega}{2\pi}A(\omega |\theta) =\frac{\tr\left[(\mathbf{S}^z_{tot})^2\right]}{\tr\left[ \mathbb{1} \right]}=\frac{N}{4},
\end{equation}
hence it makes sense to think of $A(\omega |\theta)$ (once normalized) as the conditional probability to generate an RF photon given the Hamiltonian $H(\theta)$. In that respect one might suspect that statistical measures of distance might be better suited then a simple least square error. To check the performance of each of those measures we perform a t-SNE based on each of them and look at the t-SNE loss. The idea of t-SNE is to embed high-dimensional points in low dimensions in a way that respects similarities between points, just like principal component analysis (PCA). Nearby points in the high-dimensional space correspond to nearby embedded 2-dimensional points, while distant points in the high-dimensional space are mapped to distant embedded 2D points. In general, it is impossible to faithfully represent all high-dimensional distances in low dimensions, e.g. there are many more mutually equidistant points in high-dimensions. In contrast to PCA, which simply linearly projects the data on a low dimensional hyperplane, t-SNE is designed to only care about preserving \emph{local} distances allowing distortion of large distances. This distortion partially combats the basic problem that there is simply not enough volume in low dimensions~\cite{pankaj19}. Figure~\ref{fig:DistanceSI}--A shows the distance matrix between all molecules in the dataset for the 3 different metrices under cosideration. First of all, a lot of structure if observed in all three distance metrices. While the Hellinger distance and Jensen-Shannon distance are both qualitatively and quantitiatvely similar, the Euclidean distance only captures the large distance features well. By squaring the probability distribution, the Euclidean distance effectively only cares about the mode of the distribution, supressing information about smaller peaks in the absorption spectrum.
We observe better clustering for Hellinger and Jensen-Shannon distance, this is also quantified by the increased Kullback-Leibler loss of the Euclidean t-SNE. In fact, at the level of the t-SNE loss, the Hellinger distance performs the best.

\begin{figure}[h]
	\centering
	\includegraphics[width=0.8 \textwidth]{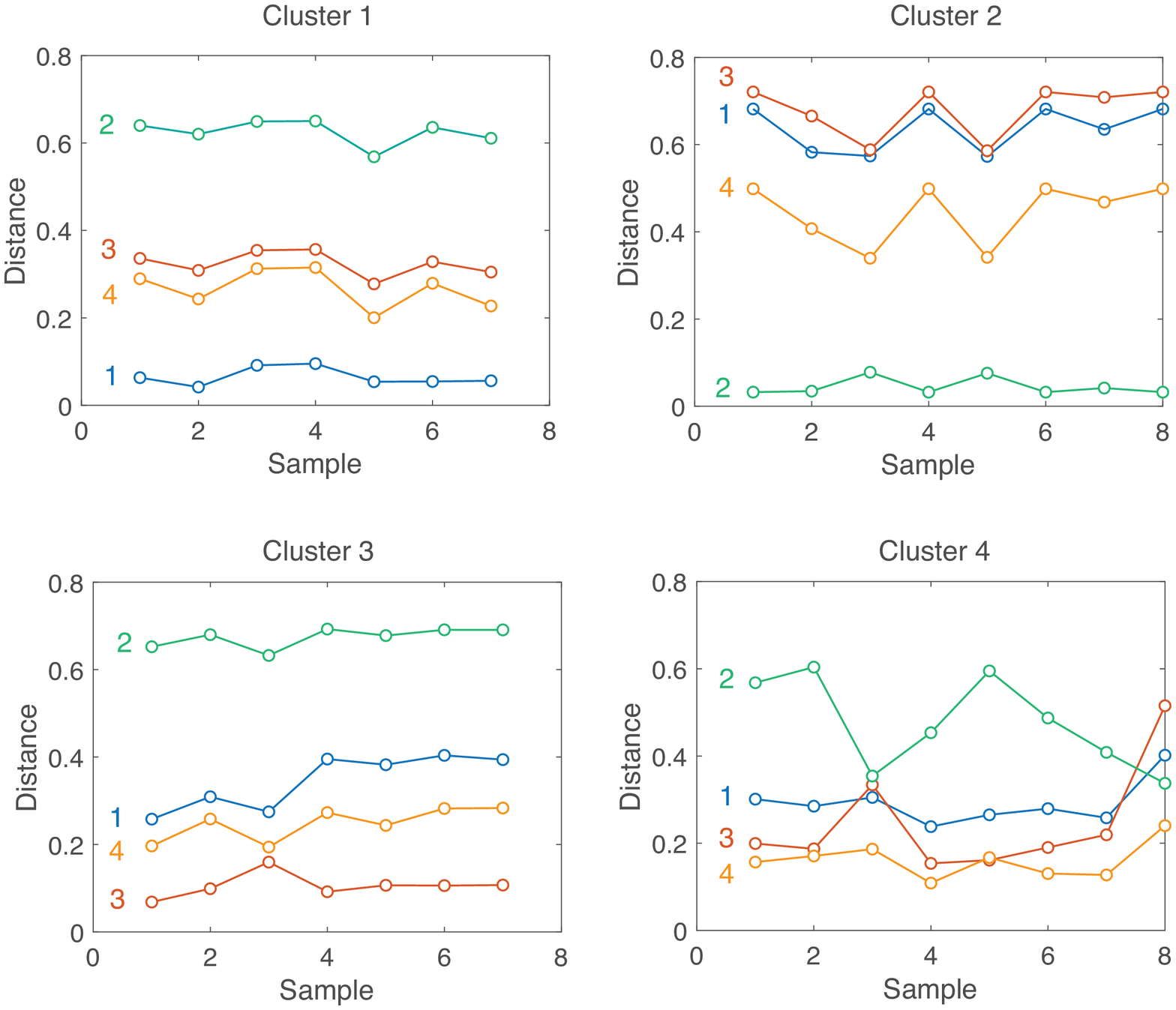}
	\caption{ \textbf{Cluster identification} 30 random samples are selected out of the set of 69. The other 39 are used to perform the cluster identification. Plots show the mean Hellinger distance between each test sample and the 39 training samples, sorted per cluster. All points are labeled according to the t-SNE clusters identified in Fig.~\ref{fig:DistanceSI}. Except for sample 5 from cluster 4, all test samples are correctly classified. }
	\label{fig:IndentifySI}
\end{figure}

\begin{figure}[h]
	\centering
	\includegraphics[width=0.4 \textwidth]{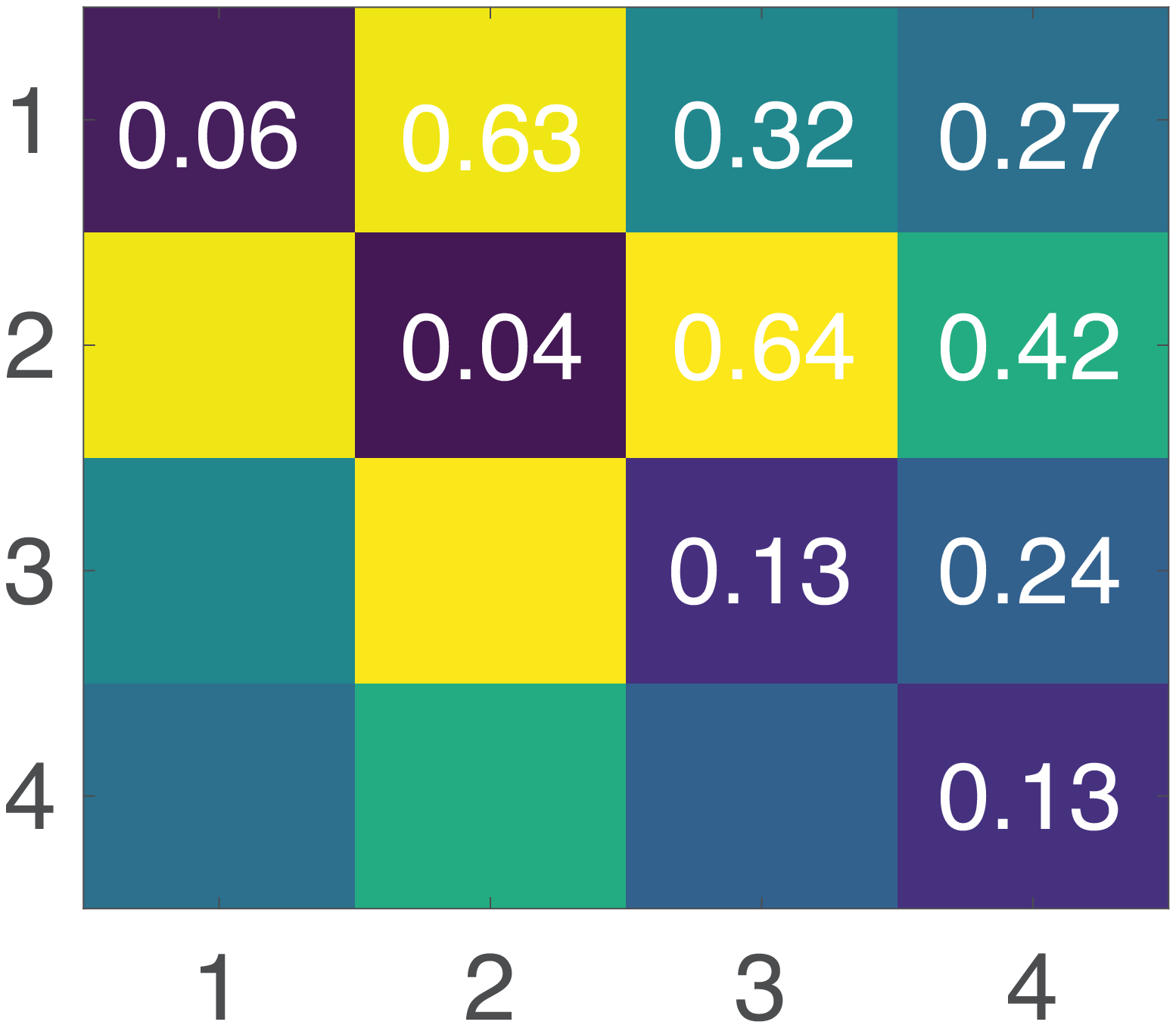}
	\caption{ \textbf{Mean Distance} Figure shows the mean Hellinger distance between samples ordered by cluster index. Clusters 1,2,3 are mutually most distinctive and mistaken most easily for cluster 4. }
	\label{fig:DistancemeanSI}
\end{figure}

To test our the robustness of the sampling procedure, we select 30 samples out of the set of 69 and check if we can classify them into the aformentioned clusters by computing the Hellinger distance to the other 39 (training) samples. The results are shown in Fig.~\ref{fig:IndentifySI}. All clusters are clearly are clearly distinguishable (~\ref{fig:DistancemeanSI}), apart from cluster 4 where one sample was incorrectly classified as cluster 3. The problem is resolved if one classifies the samples on the distance to the closest point. However, due to the smallness of the dataset it's not clear how representative the latter is. 

\clearpage

\section{\label{sec:typical} Are physical molecules typical?}
\begin{figure}[h]
	\centering
	\includegraphics[width=0.6 \textwidth]{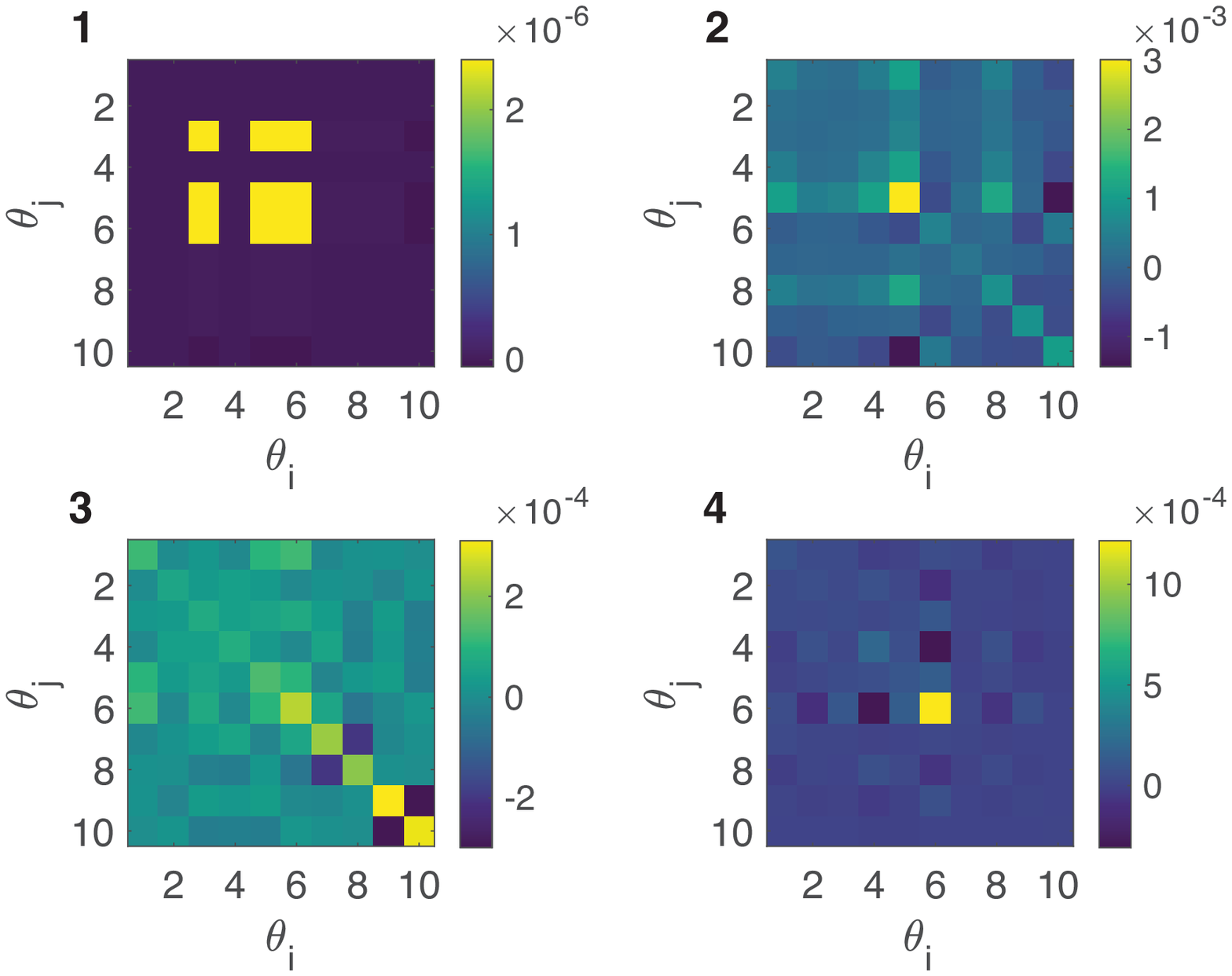}
	\caption{\textbf{FIM Features} Fisher information metric for a typical molecule out of each cluster. }
	\label{fig:Fisher}
\end{figure}

\begin{figure}[h]
	\centering
	\includegraphics[width=0.6 \textwidth]{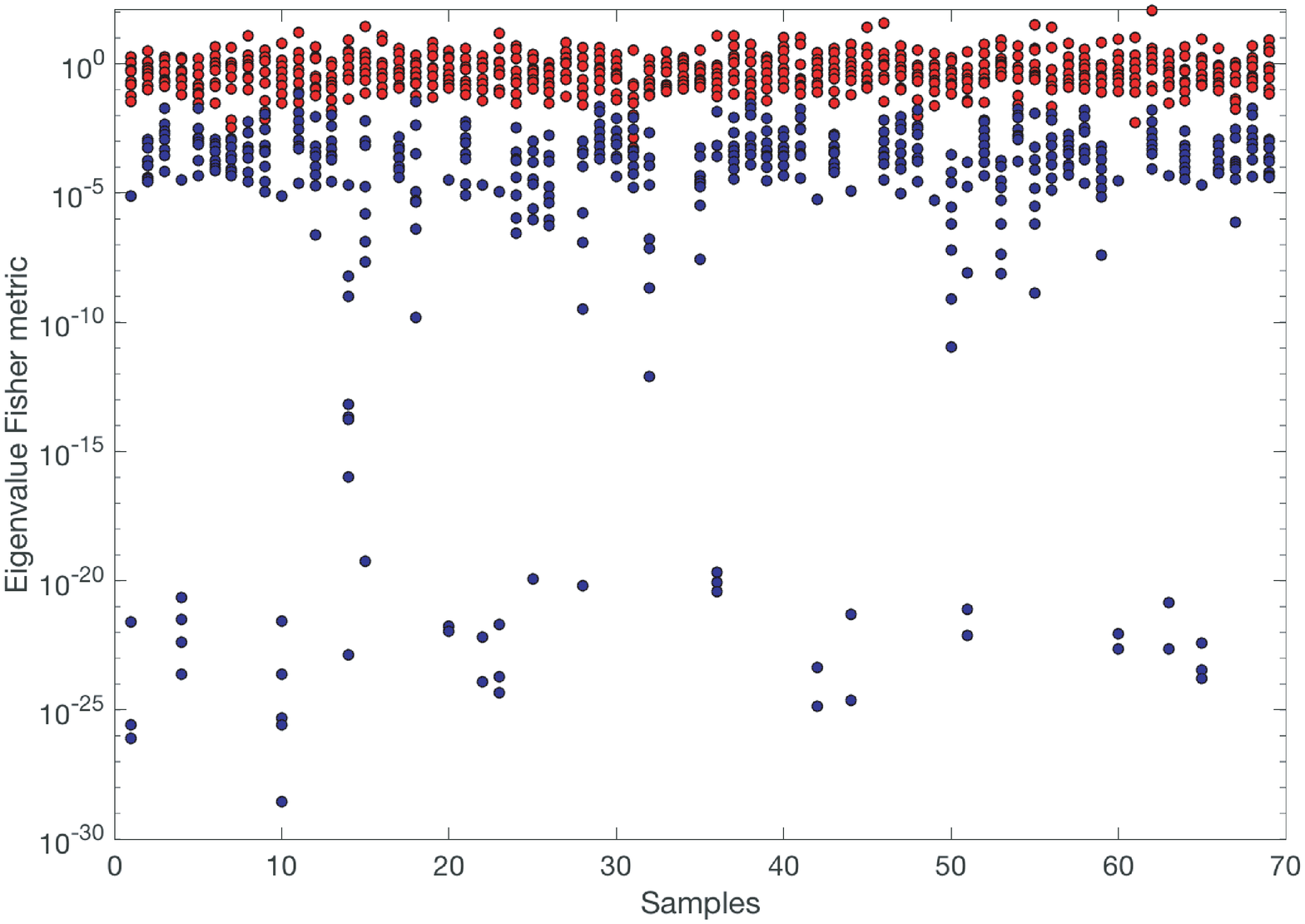}
	\caption{\textbf{FIM eigenvalues} Blue dots show the eigenvalues of the FIM for all the molecules contained in the dataset. Red dots show eigenvalues for samples obtained by sampling each of the parameters $\theta_i$ from a normal distribution with unit variance and zero mean. }
	\label{fig:FisherEig}
\end{figure}

In order to discuss whether actual spectra are atypical we need to define a notion of likelihood of a given spectra, i.e. we need a measure on the space of molecular parameters $\theta$. The measure should be unbiased by any knowledge we believe to have about physical molecules. It should only satisfy some basic consistency conditions. One very simply and natural condition is that whatever measure we are sampling from, it ought not to depend on the way we parametrize our model. That is, if one makes a change of variables $\theta'=f(\theta)$ it shouldn't change the likelihood of a given molecule since it represents exactly the same data. This parametrization invariance was first argued by Jeffreys~\cite{jeffreys46}, and it was shown that the distribution should therefore be proportional to the square root of the determinant of the Fisher information metric (FIM):
\begin{equation}
P(\theta)\propto  \sqrt{ \det I_{ij}(\theta) },
\end{equation} 
where $I_{ij}(\theta)$ is the FIM:
\begin{equation}
I_{ij}(\theta) = \int \frac{{\rm d} \omega}{2\pi} A(\omega |\theta) \frac{\partial  \log A(\omega|\theta )}{\partial \theta_i }\frac{\partial  \log A(\omega|\theta )}{\partial \theta_j }.
\end{equation}
In Bayesian inference, $P(\theta)$ is known as Jeffrey's prior and is an example of a so called uninformative prior. The question of whether molecular parameters are typical thus becomes a question about the structure of the eigenvalues of the Fisher information metric. Some representative Fisher metrices for physical molecules are shown in Fig.~\ref{fig:Fisher}.
Note that the the FIM is generally small and appears to be structured. The structure should become apperant when we look at the eigenvalues of the FIM. These are depicted in Fig.~\ref{fig:FisherEig}. Most molecules indeed seem to have a some eigenvectors -- combinations of parameters -- that are much more important then others, i.e. have eigenvalues that are exponentially larger then others. Such characteristic has been termed \emph{``sloppiness''} in the past and it has been shown to arrise naturally in multiparameter mathematical models that probe collective behavior~\cite{waterfall06,machta13}; meaning they can not probe the individual parameters but only have access to some coarse grained observable. NMR spectroscopy can be argued to be in this regime as there is no easy way to directly extract the model parameters from the spectrum. The fact that there are irrelevant combinations of parameters immediatly implies the molecules are unlikely because the determinant of the FIM must be small. In other words, these sloppy parameters represent approximate or possibly even exact symmetries of the molecules. Random models possess no symmetries and so molecules are atypical. Finally note that even the large eigenvalues of the FIM are relatively small, sampling parameters $\theta_i$ from a normal distribution with zero mean and unit variance results in significantly larger eigenvalues, see red dots in Fig.~\ref{fig:FisherEig}. In fact, the FIM eigenvalues are of $O(1)$ rather then $O(10^{-4})$.  
Given that we will only have a finite amount of data available to finally perform the inference, it would be extremely hard to converge to physical model parameters, starting from an uninformative prior. It's usefull to actually take the sloppiness of molecules into account and start from a biased prior that already takes into account the aformentioned clustering. 
\clearpage

\section{\label{sec:mix} Importance Sampling}
Recall that we aim to measure
\begin{equation}
S(t)= \frac{\tr \left[ \mathbf{S}^z_{tot}(t) \mathbf{S}^z_{tot} \right] }{\tr \left[\mathbb{1} \right]},
\end{equation}
In the diagonal basis in $\mathbf{S}^z_{tot}$, this can written as
\begin{equation}
S(t) =\sum_{i,j} m_i m_j P_t(i|j)P_0(j)=\sum_{i,j} m_i m_j P_t(i,j),
\end{equation}
with
\begin{equation}
P_t(i|j)=\vert \left< i | U_t| j \right> \vert^2 \quad {\rm and }\quad P_0(j)=\frac{1}{2^N}.
\end{equation}
At this point one can forget about quantum mechanics, it's simply hidden in the transition probability $P_t$. Here we will compare various sampling schemes, i.e. (i) uniform sampling out of $P_0$, (ii) sampling from a Gibbs distribution of the total magnetization and (iii) importance sampling. We will compare the convergence of the different estimands of $S(t)$.

\subsection*{Uniform sampling}
The most direct procedure would be to draw uniform random states $i$ -- out of $P_0$-- and propagate them under the quantum Hamiltonian, after which we measure $j$. In that case, the random variable we are estimating is $m_im_j$ and hence its variance is given by
\begin{equation}
{\rm var}[m_i m_j]= \sum_{i,j} m_i^2 m_j^2 P_t(i|j)P_0(j)-S(t)^2.
\label{eq:var0}
\end{equation}
In general, we can't compute this, as $P_t$ is hard to compute, but let's first look at $t=0$. In that case we have, 
\begin{equation}
S(t)=\frac{\tr \left[ (\mathbf{S}^z_{tot})^2 \right]}{\tr \left[ \mathbb{1} \right]}=\frac{N}{4},
\end{equation}
and 
\begin{equation}
{\rm var}[m_i m_j]= \frac{\tr \left[ (\mathbf{S}^z_{tot})^4 \right]}{\tr \left[ \mathbb{1} \right]}-S(t)^2=\frac{1}{16}(3N^2-2N)-S(t)^2=\frac{ N(N-1)}{8}.
\end{equation}
Consequently, if we want the sample variance to be constant $\epsilon^2$, such that we have a fixed precision, we need the number of samples to be of $O(N^2/\epsilon^2)$. How much will the variance increase once we evolve the system in time? Note that the random variable we are estimating is bounded, i.e. the magnetziation can not be larger than $N/2$. Hence the variance can not exceed $O(N^4)$ in general. There are however much more constraints on the present problem such that we can put a much stronger bound on the variance from expresssion~\eqref{eq:var0}. Reorganizing terms we have
\begin{eqnarray}
{\rm var}[m_i m_j] &=& \sum_{i,j} (m_j^2 \sqrt{P_t(i|j)P_0(j)}) (m_i^2 \sqrt{P_t(i|j)P_0(j)})  -S(t)^2 \nonumber \\
&\leq& \sqrt{\sum_{i,j} m_j^4 P_t(i|j)P_0(j) \sum_{i,j}  m_i^4 P_t(i|j)P_0(j)} ,
\end{eqnarray}
where we have simply used Cauchy-Schwarz and dropped the $S(t)$ term. Further note that because of the reversibility of the quantum evolution $P_t(i|j)=P_t(j|i)$, such that $\sum_i P_t(i|j)=\sum_j P_t(i|j)=1$. In addition, $P_0$ is the uniform distribution, such that we find
\begin{eqnarray}
{\rm var}[m_i m_j] & \leq& \frac{1}{2^N}\sum_{j} m_j^4=\frac{\tr \left[ (\mathbf{S}^z_{tot})^4 \right]}{\tr \left[ \mathbb{1} \right]} \leq 3 (N/4)^2,
\end{eqnarray}
Consequently, the number of samples never needs to exceed order $O(N^2/\epsilon^2)$. Finally note that, for ergodic systems, one expect the transition probability to become close to uniform at late time, i.e. $P_t(i|j)\approx 1/2^N$. The latter simply reflects the fact that those systems thermalize and effectively forget their initial conditions. In that case we can also explicitly compute the variance, 
\begin{equation}
{\rm var}[m_i m_j] =\left( \frac{\tr \left[ (\mathbf{S}^z_{tot})^2 \right]}{\tr \left[ \mathbb{1} \right]} \right)^2= (N/4)^2
\end{equation} 
Under uniform sampling we thus observe $N^2$ scaling of the variance both at early and late times, with a guarantee that it will never be larger than that.

\subsection*{Thermal sampling}
Before we go discuss how toimprove the $N^2$ in any way, let's check what happens for $t=0$ if we sample from a thermal state. Those states are particularly relevant for experiments as they might, in some particular setups, be much faster to prepare. 
\begin{equation}
\rho=\frac{1}{Z} e^{\beta \mathbf{S}^z_{tot}}. 
\end{equation}
For sufficiently small $\beta$ we have that 
\begin{equation}
S(t)\approx \beta^{-1} \tr \left[ \mathbf{S}^z_{tot}(t) \rho \right]  
\end{equation}
At $t=0$ and for $\beta \rightarrow 0$, the variance of our estimator becomes
\begin{equation}
{\rm var}[m_i]= \tr \left[  \mathbf{S}^z_{tot} \rho \right]=\frac{N}{4}.
\end{equation}
Hence the number of samples to get an precision of $\epsilon$ scales like $N/(\beta^2 \epsilon^2)$. If one demands all subleading terms in the Taylor expansion of $\rho$ to be subleading, one should set $\beta \sim 1/N$. Since we want to get $S(t)$ at precision $\epsilon$ anyway, one needs to set $\beta =O( \sqrt{\epsilon} /N )$. Indeed, expanding $\rho$ in powers of $\beta$, one gets
\begin{equation}
S(0) =\beta^{-1} \tr \left[ \mathbf{S}^z_{tot} \rho \right]  \propto \tr \left[ ( \mathbf{S}^z_{tot})^2 \right] + \frac{\beta^2}{6} \tr \left[ (\mathbf{S}^z_{tot})^4 \right] +O(\beta^4 N^3).
\end{equation}
The second term is of $O(\beta^2 N^2)$, implying we need to have $\beta^2 < \epsilon/N^2$ to achieve an accuracy of $\epsilon$. Combined with the scaling of the variance, we get a scaling of $O(N^3/\epsilon^3)$ to reach the desired accuracy and precision.

\subsection*{Importance sampling}
Sampling from the thermal state is thus a factor $N/\epsilon$ less efficient as uniform sampling. One could wonder whether there is a distribution such that we would need less than $O(N^2)$ samples. One can recast the problem of estimating $S(t)$ as 
\begin{equation}
S(t) =\sum_{i,j} \frac{P_0(j)m_i m_j}{Q_0(j)} P_t(i|j)Q_0(j),
\end{equation}
where $Q_0$ is the distribution from which we will sample initial states. This clearly gives the same correlation function, but the stochastic variable $r$ we are estimating is now different, i.e. 
\begin{equation}
r=\frac{P_0(j)m_i m_j}{Q_0(j)}.
\end{equation}
The variance now becomes 
\begin{equation}
{\rm var}[r]=\sum_{i,j} \left[ \left( \frac{P_0(j)m_i m_j}{Q_0(j)} \right)^2 P_t(i|j)Q_0(j) \right]-S(t)^2
\end{equation}
We'd have an optimal sampling algorithm (at least in the central limit regime), if we minimize the variance of the estimator with respect to our sampling distribution. Hence deriving the variance with respect to the sampling distribution $Q_0$, and putting in a Lagrange multiplier to keep the distribution normalized we find
\begin{equation}
\sum_i  \left( \frac{P_0(j)m_i m_j}{Q_0(j)} \right)^2 P_t(i|j)=\mu, 
\end{equation}
hence 
\begin{equation}
Q_0(j) =\frac{1}{\mu(t)} |m_j|  P_0(j)\sqrt{\sum_i m_i^2 P_t(i|j)} \quad {\rm with}\quad \mu(t)= \sum_j  |m_j|  P_0(j)\sqrt{\sum_i m_i^2 P_t(i|j)} .
\end{equation}
The variance then becomes 
\begin{equation}
{\rm var}[r]=\mu(t)^2-S(t)^2
\end{equation}
In general the optimal distribution depends on time through the transition probability $P_t$. Since we have no access to that distribution we can't perform optimal sampling. However, as discussed before there are two limits worth investigating $t=0$ and $t \rightarrow \infty$. When $t=0$, $P_t=\delta_{ij}$, consequently
\begin{equation}
Q_0=\frac{N}{4} \frac{m_j^2}{2^N}, \quad {\rm and} \quad {\rm var}[r]=0
\end{equation}
This makes sense, since the random variable we are estimating $r=N$ is just a constant. Hence there is nothing to estimate. Note that at late times, when $P_t\approx 1/2^N$, this sampling distribution results in a variance of $(N/4)^2$, which is identical to the late time variance of the uniform sampling problem. In fact the variance at all times is
\begin{equation}
{\rm var}[r] = \left(\frac{N}{4}\right)^2-S(t)^2.
\end{equation} 
Improving on the uniform sampling by a factor $3$. 
Finally note that at late times the optimal sampling distribution should tend to 
\begin{equation}
Q_0 \propto |m_j|,
\end{equation}
which would result in a variance ${\rm var}[r] \approx N^2/8\pi$ in the large-N limit. Consequently, it only reduces the variance of the estimand at late times by a factor $\pi/2$ over the other sampling schemes, while significantly increasing the short time fluctuations. In conclusion it thus seems most efficient to sample from the short time optimal distribution, as it supresses the variance to zero at early times while always outperforming uniform sampling.

\end{document}